\documentclass[fleqn,usenatbib]{mnras}

\usepackage[T1]{fontenc}
\usepackage{ae,aecompl}

\usepackage{graphicx}	
\usepackage{amsmath}	
\usepackage{amssymb}	
\usepackage{multicol}
\usepackage{xspace}
\usepackage{xcolor}
\usepackage{booktabs}
\usepackage[normalem]{ulem}
\usepackage{newtxtext,newtxmath}

\def\sec#1{Sec.~\ref{sec:#1}}
\def\app#1{App.~\ref{sec:#1}}
\def\Fig#1{Fig.~\ref{fig:#1}}
\def\Table#1{Table~\ref{tab:#1}}
\def\Eq#1{Eq.~(\ref{eq:#1})}
\newcommand{\nlg}{NIHAO-LG\xspace}
\newcommand{\lgn}{NIHAO-LG(nmd)\xspace}
\newcommand{\rz}{\,$z=0$\xspace}

\def \hi{\ion{H}{I}\xspace}

\title[NIHAO-LG]{NIHAO-LG: the uniqueness of Local Group dwarf galaxies}

\author[N. Arora et al.]
{Nikhil Arora,$^{1}$\thanks{E-mail: nikhil.arora@queensu.ca}
Andrea V. Macci\`o,$^{2,3,4}$
St\'ephane Courteau,$^{1}$
Tobias Buck,$^{5}$
\newauthor
Noam I. Libeskind,$^{5,6}$
Jenny G. Sorce,$^{5,7,8}$
Chris B. Brook,$^{12,13}$
Yehuda Hoffman$^{11}$
\newauthor
Gustavo Yepes,$^{9,10}$
Edoardo Carlesi$^{11}$
and Connor Stone$^{1}$
\\
$^{1}$Department of Physics, Engineering Physics and Astronomy, Queen’s University, Kingston, ON K7L 3N6, Canada\\
$^{2}$New York University Abu Dhabi, PO Box 129188, Abu Dhabi, United Arab Emirates\\
$^3$Center for Astro, Particle and Planetary Physics (CAP$^3$), New York University Abu Dhabi\\
$^{4}$Max-Planck-Institut für Astronomie, Königstuhl 17, D-69117 Heidelberg, Germany\\
$^{5}$ Leibniz-Institut für Astrophysik Potsdam (AIP), An der Sternwarte 16, D-14482 Potsdam, Germany\\
$^{6}$ University Lyon, University Claude Bernard Lyon 1, CNRS, IP2I Lyon/IN2P3, IMR 5822, F-69622, Villeurbanne, France\\
$^{7}$Univ. Lyon, ENS de Lyon, Univ. Lyon1, CNRS, Centre de Recherche Astrophysique de Lyon UMR5574, F-69007, Lyon, France\\
${^8}$Université Paris-Saclay, CNRS, Institut d'Astrophysique Spatiale, 91405, Orsay, France\\
$^{9}$Departamento de F\'isica Te\'{o}rica, M\'{o}dulo 8, Facultad de Ciencias, Universidad Aut\'{o}noma de Madrid, 28049 Madrid, Spain\\
$^{10}$Centro de Investigaci\'{o}n Avanzada en F\'{\i}sica Fundamental (CIAFF), Universidad Aut\'{o}noma de Madrid, 28049 Madrid, Spain\\
$^{11}$Racah Institute of Physics, Hebrew University, Jerusalem 91904, Israel\\
$^{12}$Universidad de La Laguna. Avda. Astrofísico Fco. Sánchez, E-38200, La Laguna, Tenerife, Spain\\
$^{13}$Instituto de Astrofísica de Canarias, Calle Via Láctea s/n, E-38206 La Laguna, Tenerife, Spain\\
}

\date{Accepted XXX. Received YYY; in original form ZZZ}

\pubyear{2022}

\begin{document}
\label{firstpage}
\pagerange{\pageref{firstpage}--\pageref{lastpage}}
\maketitle
 
\begin{abstract}
   Recent observational and theoretical studies of the Local Group (LG) dwarf galaxies have highlighted their unique star-formation history, stellar metallicity, gas content, and kinematics. 
   We investigate the commonality of these features by comparing constrained LG and field central dwarf halo simulations in the Numerical Investigation of a Hundred Astrophysical Objects (NIHAO) project. 
   Our simulations, performed with NIHAO-like hydrodynamics which track the evolution of the Milky Way (MW) and M31 along with $\sim$100 dwarfs in the LG, reveal the total gas mass and stellar properties (velocity dispersion, evolution history, etc.) of present-day LG dwarfs to be similar to field systems. 
   However, relative to field galaxies, LG dwarfs have more cold gas in their central parts and more metal-rich gas in the halo stemming from interactions with other dwarfs living in a high-density environment like the LG. 
   Interestingly, the direct impact of massive MW/M31 analogues on the metallicity evolution of LG dwarfs is minimal; LG dwarfs accrete high-metallicity gas mostly from other dwarfs at late times. 
   We have also tested for the impact of metal diffusion on the chemical evolution of LG dwarfs, and found that it does not affect the stellar or gaseous content of LG dwarfs. 
   Our simulations suggest that the stellar components of LG dwarfs offer a unique and unbiased local laboratory for galaxy-formation tests and comparisons, especially against the overall dwarf population in the Universe.
\end{abstract}




\section{Introduction}

In the $\Lambda$ cold dark matter ($\Lambda$CDM) paradigm, galaxy growth occurs through hierarchical assembly and secular evolution \citep{White1978, White1991, Kormendy1992}. 
As a result, galaxies are dynamic entities that form and assemble into groups and clusters.
This holds for our Milky Way (MW) and the Andromeda galaxy (M31) which, along with numerous dwarf galaxies, form the Local Group (LG) of galaxies \citep{Mateo1998, Mcconnachie2012, Kirby2013}.

LG dwarf galaxies offer a superb laboratory to study the physics of galaxy formation and evolution. 
Indeed, numerous observational, theoretical, and numerical studies of the very nearby dwarf galaxies have been conducted \citep[][to name a few]{Mateo1998, Mcconnachie2012, Kirby2013, benitez2015, benitez2016, Libeskind2020}. 
A fundamental question arises regarding LG dwarfs as representative of the general dwarf galaxy population in the Universe. 
Alternatively one can ask whether the study of the LG dwarfs teaches us about galaxy formation at large, in the field. 
One might think that tidal interactions between dwarf galaxies and MW/M31 could lead to formation of streams and eventual phase mixing into the stellar haloes of MW/M31. 
These interactions can also transfer pristine gas from MW/M31 to surrounding dwarfs which may induce star formation and evolution of metals in dwarf galaxies \citep{Buck2020a}. 
Such interactions could also lead to the stripping of gas leading to a “freeze out" of the stellar population in LG dwarfs. 
Could such environmental interactions distinguish the LG dwarfs from the general population of dwarfs in the Universe? 
What are the cosmological ramifications of such differences? 
This Copernican question has directly motivated the present paper.

The stellar assembly of dwarf galaxies in the LG is sensitive to internal (feedback and winds) and environmental (ram pressure and tidal stripping) processes, as well as the reionization of gas due to the ultraviolet (UV) background at early times.
As a result, the star-formation histories (SFHs) of many LG dwarfs have been extensively studied to better understand the influence of these internal and external processes. 
Some of these dwarfs form a large fraction of their stars at early times and are quenched after reionization \citep{Sand2010, Brown2012, Okamoto2012}, while others show moderate and continuous star formation to the present day \citep{Weisz2014, Gallart2015}.
Overall, SFHs of dwarf galaxies are the product of reionization and the environment in which they are formed.
\cite{Gallart2015} have also shown that dwarfs with no current star formation could form in dense, cluster/group-like environments.
Dwarfs that form in low-density environments could thus retain their gas reservoir to fuel continuous star formation and younger stellar populations.

The presence of MW/M31 is likely to influence the gas, and therefore the stellar, content of LG dwarfs.
Indeed, the fraction of neutral hydrogen in LG dwarf galaxies increases with distance from the MW \citep{Spekkens2014, Putman2021}.
Dwarfs within the virial radius of the MW are especially deficient in cold gas while, in comparison, systems outside the virial radius have 100 times higher neutral gas fractions \citep{Einasto1974,Grcevich2009,Mcconnachie2012,Spekkens2014}.
Gas-poor systems result largely from interactions with the hot halo of the MW typically due to ram pressure stripping, viscous stripping, and starvation \citep{Gunn1972,Hester2006,Kawata2008,Fillingham2016}.

While the evolution of dwarf galaxies seems broadly understood, a key question remains: Are the observed features of LG dwarfs unique or do field dwarfs show similar evolutionary tracks?  
A robust answer to this question must rest on two pillars. 
One is our ability to simulate the formation of the LG and its nearby neighbourhood in way that reproduces the main features of the LG. 
Namely, simulations that are constrained to emulate the formation and evolution of the LG dwarfs within the ‘context’ of the actual observed LG. 
The other pillar consists of a detailed census of LG and field dwarfs in the local Universe.
Depending on the choice of Stellar Mass $--$ Halo Mass relation (SHMR), especially for the low-mass end, about $0.3-0.4~deg^{-2}$ of field dwarfs with surface brightness of $\sim$30 mag arcsec$^{-2}$ are expected in the local volume, with distances between 3 and 10 Mpc from us \citep{Danieli2018}.
Indeed the LG includes approximately 100 dwarf galaxies with a magnitude range $-17\geq M_{\rm V}\geq-7$ within a volume of 3 Mpc \citep{Mcconnachie2012}.
Until the advent of deep large sky surveys, such as those provided by the Rubin Observatory with a limiting magnitude of 32 mag arcsec$^{-2}$, comparisons with complete census of field dwarfs is beyond reach. 
While the presence (or lack) of unique aspects of the LG cannot currently be characterized with observed galaxies, high-resolution simulations may provide much valuable insight.

Identifying differences between LG and field dwarf galaxies can help in better isolating the role of environment in shaping the galaxy properties in the local Universe.
In the context of other groups and clusters, galaxy properties like mean stellar age \citep{Thomas2005, Clemens2006}, morphology \citep[][and references therein]{Blanton2009}, colour \citep{Wilman2010, Cluver2020} and star formation rates \citep[SFR;][]{Fossati2015} have been shown to depend strongly on environment.
Similar conclusions have been drawn about the influence of MW/M31 on LG satellite dwarfs (e.g., \citealt{CLUES2010,benitez2015,benitez2016, Buck2019, Genina2019, Libeskind2020, DiCintio2021}).
However, if similarities exist between LG and field dwarf, nearby population of dwarf galaxies (which are more easily accessible) can be used as a proxy for distant systems.

This paper presents such a comparison between the dwarf populations found in the LG and in the field through high-resolution simulations in order to highlight any similarities and/or differences.
To this end, we take advantage of the Numerical Investigation of a Hundred Astrophysical Objects (NIHAO) simulations which trace the evolution of individual dark matter haloes and their baryonic components for the complete history of the Universe \citep{nihao_main}.
The NIHAO simulations have already proven to be successful in matching various observations aspects of galaxy formation and evolution \citep[][and references therein]{Maccio2016, Maccio2017, Obreja2019, Buck2020b, Blank2021}.
More details about these simulations are provided in \sec{nihao}. 

Our field dwarf sample uses NIHAO haloes as presented in \cite{nihao_main}. 
The LG dwarf sample is comprised of two constrained LG simulations with initial conditions provided from the CLUES collaboration \citep{CLUES2010}, presented here for the first time, and performed with the exact same code and galaxy formation model used for the NIHAO project.
The constrained LG simulation allowed for the evolution of LG environment containing the MW- and M31-like haloes and their associated dwarfs. 
Our comparisons rely mostly on common galaxy scaling relations such as the stellar mass$-$metallicity relation (hereafter MZR; \citealt{Gallazzi2005, Mcconnachie2012, Kirby2013}), 
stellar mass--gas mass relation \citep{Peeples2014}, , 
and the SHMR \citep{Behroozi2013, Moster2014}.

Because the formation and evolution of dwarfs in high-resolution simulations is sensitive to the implemented subgrid prescriptions such as feedback, chemical enrichment, and metal diffusion, 
we have also compared the subgrid physics formalism that may drive the pre-enrichment of LG dwarfs; namely metal diffusion.
The impact of metal diffusion on the stellar assembly, SFRs, and chemical evolution of dwarf galaxies remains unclear. 
For instance, while \cite{Su2017} used the FIRE (Feedback In Realistic Environment; \citealt{Hopkins2018}) simulations to show that the subgrid metal diffusion does not impact SFRs on the galactic scales,
\cite{Pilkington2012} and \cite{Williamson2016} found opposite results for the abundance of low-metallicity stars in dwarf galaxies (see also \citealt{Kawata2014} and \citealt{Escala2018}). 
Given the current muddled picture about the evolution and abundance of metals in LG and field dwarfs, we present results from two constrained LG simulations; with and without metal diffusion.

This paper is organised as follows: \sec{sims} outlines the NIHAO galaxy formation simulation which provides the field galaxy sample for this study, and presents the two constrained LG simulations. 
\sec{lg_gal} highlights the similarities and differences between the constrained LG dwarf galaxies and the NIHAO dwarf galaxies at $\rm z\sim 0$.
A multifaceted analysis of the dwarf galaxies using various galaxy scaling relations is also presented. 
The evolution of galaxy properties for the NIHAO and NIHAO-LG simulations to better understand the difference seen at \rz is addressed in \sec{evol}, and conclusions are presented in
\sec{conclusion} as we ponder the uniqueness (or lack thereof) of LG dwarfs and its implications. 

\section{Simulations}
\label{sec:sims}
\subsection{NIHAO galaxy formation simulations}
\label{sec:nihao}

Our field central galaxy sample relies on the 'NIHAO' cosmological zoom-in simulations presented in \cite{nihao_main}.  
The simulations were run with a flat $\Lambda$CDM cosmology with parameters from the \textit{Planck satellite}: $H_0 = 100h{\rm \,km\,s^{-1}\,Mpc^{-1}}$ with $h =0.671,\, \Omega_{\rm m}=0.3175,\, \Omega_{\rm \Lambda}=0.6824,\, \Omega_{\rm b}=0.049,\, \sigma_{\rm 8}=0.8344$ and $n=0.9624$ \citep{planck14}.
The hydrodynamics were performed with the updated N-body SPH solver {\scriptsize GASOLINE2} \citep{Wadsley2017} which includes the treatment of $\rm P/\rho^2$ proposed by \cite{Ritchie2001}. 
Gas cooling was performed through hydrogen, helium and various metal-lines in a uniform UV ionizing background. 
Photoionization and heating of the gas also occur via UV background and Compton cooling with temperatures from 10 to $\rm 10^9\,{\rm K}$ \citep{Shen2010}.

All NIHAO galaxies were allowed to form stars provided that the gas follows the Kennicutt-Schmidt law \citep{Kennicutt1998} with suitable density and temperature thresholds, $\rm T<15000\,{\rm K}$ and $\rm n_{\rm th}>10.3\,{\rm cm^{-3}}$. 
Energy is re-injected back into the interstellar medium (ISM) from stars through stellar and blast wave supernova feedback. 
Massive stars also ionize the ISM prior to their supernova explosion; this is referred to as ``early stellar feedback" \citep[ESF][]{Stinson2006, nihao_main} where 13 per\,cent of the total stellar flux of $2\times 10^{50}{\rm erg\,M_{\odot}^{-1}}$ is injected into the ISM. 
This differs from the original prescription presented in \cite{Stinson2013} in order to account for the increased mixing of gas and aligns with the abundance matching results on MW scale  \citep{Behroozi2013}. 
For supernova feedback, massive stars with $8\,{\rm M_{\odot}}<M_{*}<40\,{\rm M_{\odot}}$ inject energy of $10^{51}{\rm erg}$ and metals into the the ISM.
Because the energy is injected into high density gas, it radiates away via efficient cooling on short time-scales. 
Therefore, for gas particles inside the blast radius, cooling is delayed by $30\,{\rm Myr}$ \citep{Stinson2013} to prevent immediate radiation from high density gas particles \citep{Stinson2006}. 
The spiral and dwarf galaxies generated by NIHAO simulations have been shown to match numerous observed galaxy properties and scaling relations \citep{Maccio2016,Obreja2016,Buck2017,Dutton2017}. 
The NIHAO galaxy simulations result in numerically converged galaxies as shown by the ultra high-definition NIHAO runs \citep{Buck2020b}.

\subsection{NIHAO Local Group simulations}

\begin{table*}
\caption{Various properties of our three NIHAO simulations. Columns give the box size, mass resolution of the DM particles, softening length of the dark and gas particles, number of haloes at \rz, usage of metal diffusion, and the simulated environment.}
\centering
\begin{tabular}{@{}ccccccccc@{}}
\toprule
Simulation     & Box Size $\rm [Mpc\,h^{-1}]$ & $\rm m_{dark}\,[M_{\odot}]$ & $\rm \epsilon_{dark}\,[pc]$ & $\rm \epsilon_{gas}\,[pc]$ & $\rm N_{haloes}$ & $\rm N_{dwarfs}$ & Metal Diffusion? & Environment\\
\midrule
NIHAO          & 60.1   & $1.74\times 10^{6}$      & 931.5      & 398.0    & 91        & 37 & Yes   & Field       \\
NIHAO-LG       & 100.0  & $1.62\times 10^{6}$      & 860.3      & 487.7    & 104       & 64 & Yes   & Local Group \\
NIHAO-LG (nmd) & 100.0  & $1.62\times 10^{6}$      & 860.3      & 487.7    & 115       & 73 & No    & Local Group \\ \bottomrule
\end{tabular}
\label{tab:sims}
\end{table*}

Our sample of dwarf galaxies in a LG environment was created with initial conditions from the Constrained Local UniversE Simulations (CLUES) project\footnote{http://www.clues-project.org.} \citep{CLUES2010, Carlesi2016, Sorce2016b, Libeskind2020}.
Constrained simulations allowed us to track the position and environment of the MW, M31-like galaxies and their associated dwarfs.
The haloes were identified and tracked in a cosmological box of $100\,{\rm Mpc\,h^{-1}}$ on a side, and constrained by observational data of the nearby Universe.
This resulted in a high-resolution simulated spherical region of approximately 5\,Mpc in radius.
The initial conditions for the simulation box relied on the Wiener filter \citep[WF,][]{Hoffman1991, Hoffman2009}, a Bayesian linear algorithm, and constrained realizations of the Gaussian matter density field from observations of the local Universe and an assumed prior model \citep{Zaroubi1995}.
The WF allowed for the construction of the cosmic displacement field needed to robustly model particle positions as a function of time for the constrained objects. 
The cosmic displacement field was created via a peculiar velocity field of the local Universe using the CosmicFlows-2 (CF2) catalogue of galaxy redshift and direct distances \citep{Tully2013}.
Malmquist biases and lognormal errors in the CF2 data set are corrected using the bias minimization technique described in \cite{Sorce2015}.
To correct for such displacements due to cosmic evolution and to calculate the positions of the galaxies progenitors, the reverse Zel'dovich approximation was applied \citep{Doumler2013, Sorce2014}.
Finally, renormalization of the velocity field was performed to get particles with initial velocity values of the particles in the simulation box.
A more detailed description of the CLUES initial conditions is found in \cite{Carlesi2016, Sorce2016b, Sorce2018} and \cite{Libeskind2020}.

The initial conditions from CLUES simulations with the cosmology of NIHAO and hydrodynamics from {\scriptsize ESF-GASOLINE2} lead to the constrained NIHAO-LG simulations.
Along with the effect of environment (field vs. LG), we also monitored the variations in galaxy properties due to changes in subgrid physics implementations; especially the chemical evolution in dwarfs.
To that end, we created two sets of constrained LG simulations: (i) \nlg,: full NIHAO cosmology and hydrodynamics included along with metal diffusion from \cite{Wadsley2008}, and (ii) \lgn,: full NIHAO cosmology and hydrodynamics with no metal diffusion.
We used the Chabrier initial mass function \citep{Chabrier2003} for stellar sampling in the simulation.
As we sought differences between field central dwarfs and LG dwarfs, the variation of chemical evolution was an obvious implementation to modulate.
\Table{sims} summarizes the basic properties of the three NIHAO simulations used in this study.

\subsubsection{Dwarfs in the LG simulations}

\begin{figure}
    \centering
    \includegraphics[width=\linewidth]{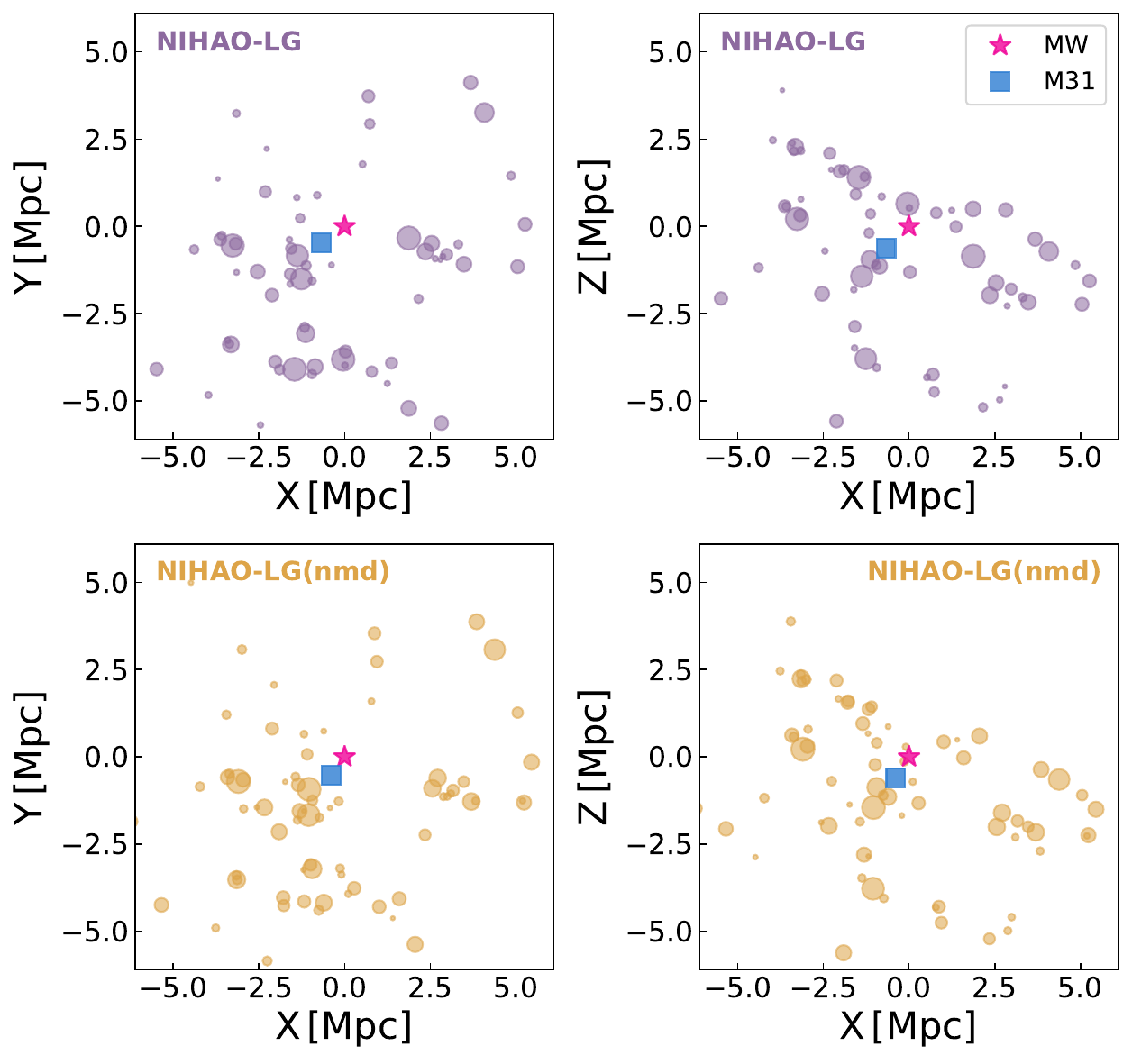}
    \caption{Distribution of dwarf galaxies ($7.0\leq\log(M_*/M_{\odot})\leq 9.5$) within our constrained LG simulations at \rz. 
    The top and bottom panels represent NIHAO-LG and NIHAO-LG(nmd), while the left- and right-hand panels show the face-on and edge-on orientations. 
    Both LG simulations are centered centred on the MW (pink star) and M31 (blue square) analogues.
    In all panels, the dwarf systems are shown as circles whose sizes scale with stellar mass at \rz.}
    \label{fig:LG_pos}
\end{figure}

From the constrained LG simulations, we selected all haloes with 100 per cent high-resolution DM particles (i.e., no pollution from low-resolution particles), having at least 1000 particles (baryon and DM) and at least 100 stellar particles.
Our dwarf galaxies are defined as central systems at \rz with $7.0\leq\log(M_*/M_{\odot})\leq 9.5$ from all pure haloes in the constrained LG simulations. 
Unless stated otherwise, all quantities (stellar mass, cold gas mass, etc.) were measured within $0.2R_{\rm 200}$; where $R_{\rm 200}$ is the radius within the average density is 200 times the critical matter density of the Universe; $\rho_{\rm crit,0}=3H_{0}^2/8\pi G$.
We expect numerical resolution and convergence to only play a small role in the results presented here. 
The hydrodynamics, performed with \textsc{gasoline2}, as is the case for NIHAO and the constrained LG simulations, were stable with respect to resolution \citep{Maccio2017, Maccio2019}.
Galaxy properties in the NIHAO simulations have also been shown to converge despite the varying spatial and mass resolution over a large range of halo masses.
The highest resolution NIHAO dwarf galaxies presented in \cite{Maccio2017} yielded galaxy properties in broad agreement with the general NIHAO-simulated galaxy population \citep{nihao_main}.

The distribution of our LG simulations (with and without metal diffusion) is presented in \Fig{LG_pos} through face-on and edge-on views centred on the respective MW analogue.
MW and M31 analogues for both constrained LG simulations are shown as a pink star and a blue square respectively.
The size of the dwarf galaxies (shown as circles) scales with their respective stellar masses at \rz. 
Both LG simulations, NIHAO-LG and NIHAO-LG(nmd), resulted in similar spatial distributions of the central dwarf systems extending out to $\sim$5\,Mpc.
In each constrained LG simulation, two massive haloes were found with a total mass $M_{\rm 200}\sim 10^{12}{\,\rm M_{\odot}}$, a ratio of stellar masses for the massive haloes of $M_{*}^{\rm MW}/M_{*}^{\rm M31}\sim 1.05$, and circular velocities of $\sim$230\,$\rm km\,s^{-1}$.
Both massive haloes have virial radii of $\sim$ 200\,kpc and are separated by $\sim$ 1\,Mpc.
Altogether, both NIHAO-LG simulations yielded similar stellar and cold gas distribution for the dwarfs halos selected for this study.

We also investigated NIHAO-LG and NIHAO-LG(nmd) for any ``backsplash'' dwarf halos; i.e., whether LG dwarfs were ever a satellite of the MW and/or M31 analogues \citep{Buck2019}.
However, within the stellar mass range of $7.0\geq\log(M_{\rm *}/M_{\odot})\geq 9.5$, no backsplash haloes were found in our constrained LG simulations. 
Only a few of backsplash haloes exist for central LG dwarfs with stellar mass of $\log(M_{\rm *}/M_{\odot})<7.0$.
The differences between LG and field central dwarf galaxy properties, discussed later in this paper, are attributed mainly to the global LG environment with possible relatively negligible contributions from the two massive haloes, MW/M31.

\subsubsection{Observational comparisons}
\begin{figure*}
    \centering
    \includegraphics[width=\linewidth]{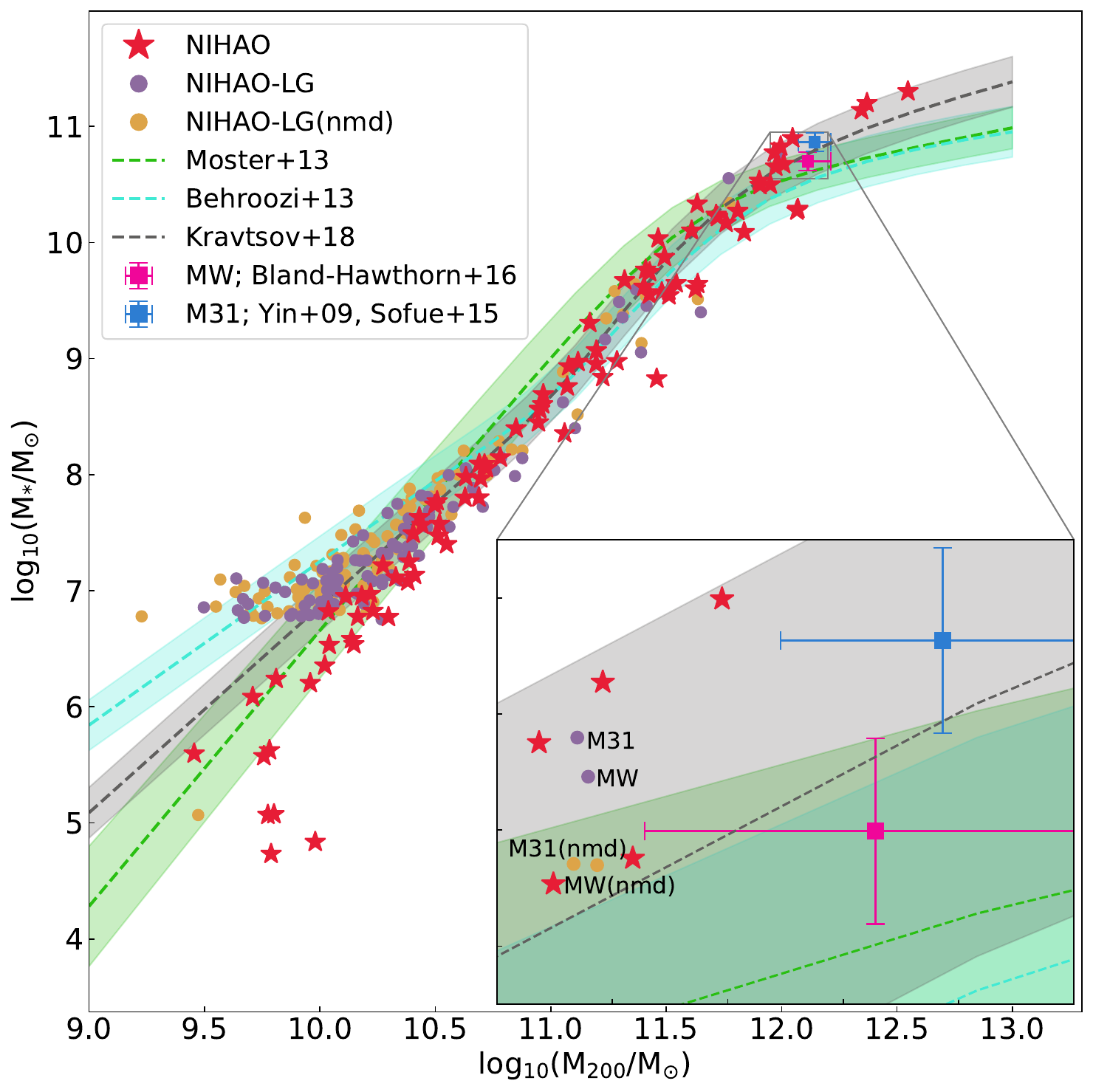}
    \caption{SHMR at redshift $z=0$ ffor NIHAO (red stars) and two constrained LG simulations with (purple circles) and without (gold circles) metal diffusion. 
    The dashed lines and shaded regions depict popular abundance matching relations from \protect\cite{Moster2014}, \protect\cite{Behroozi2013} and \protect\cite{Kravtsov2018}. 
    The zoomed inset panel show MW and M31 analogues and comparisons with massive NIHAO spirals. 
    The observed stellar and halo masses for the MW \citep{Bland-Hawthorn2016}, pink square, and M31 (stellar mass; \citealt{Yin2009} and halo mass; \citealt{Sofue2015}), blue square, are also presented.}
    \label{fig:stellarhalo}
\end{figure*}

\Fig{stellarhalo} shows the SHMR at \rz for the NIHAO and constrained NIHAO-LG simulations.
For comparison, the halo abundance matching relations from \cite{Moster2014}, \cite{Behroozi2013}, and \cite{Kravtsov2018} are also shown. 
The inset panel in \Fig{stellarhalo} features the NIHAO field and constrained LG analogues for the MW and M31 along with observed measurements.
The simulated stellar and halo masses for the MW/M31 agree with current estimates of the observed stellar and halo masses for MW \citep{Bland-Hawthorn2016} and M31 \citep{Yin2009, Sofue2015}.
As stated in \cite{nihao_main}, the NIHAO galaxies, which serve as our field galaxy sample, also agree well with halo abundance matching results \citep{Kravtsov2018}. 
The dwarf galaxies ($\log(M_{\rm 200}/M_{\odot}) \leq 10.5$) in the constrained LG simulations also match the abundance matching relation from \cite{Behroozi2013}.
Likewise, the galaxies from the constrained LG simulations with $ \log(M_{\rm 200}/M_{\odot}) \geq 10.5$ match the SHMR of NIHAO field centrals and the relation presented in \cite{Kravtsov2018}.
However, dwarfs in both LG simulations have stellar masses that exceed their NIHAO field counterparts by $0.2-0.3\,{\rm dex}$.

The shallow potential of LG dwarf galaxies is insufficient to retain high-metallicity gas due to strong supernova feedback and stellar winds \citep{Dekel1986}. 
The LG dwarfs may also have lower DM fraction for the same stellar mass relative to field systems as a result of interaction episodes with host halos and other systems \citep{Buck2019}.

\begin{figure*}
    \includegraphics[width=\linewidth]{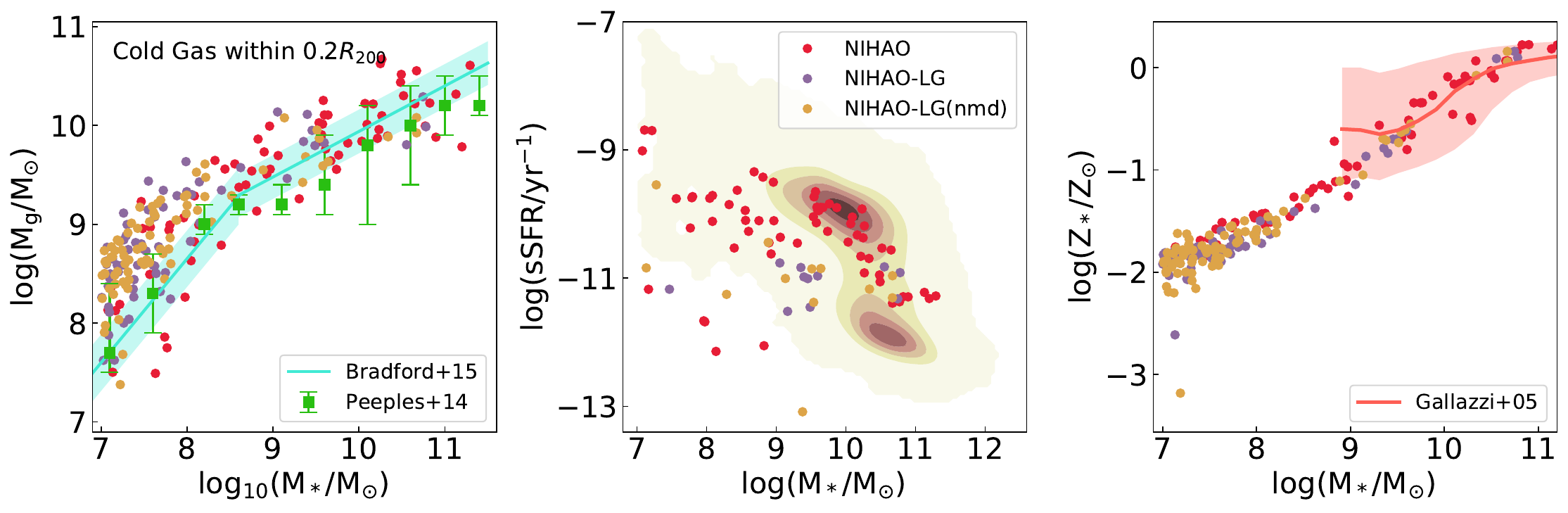}
    \caption{Various baryonic scaling relations at present day  (\rz) for NIHAO-LG simulations and comparisons with observations.
    Left-hand panel: Cold gas mass ($T<20000\,{\rm K}$) vesus stellar mass at $z=0$ for NIHAO, NIHAO-LG, and NIHAO-LG(nmd) simulated galaxies presented as red, purple, and gold circles. Observed cold gas mass-stellar mass relations from \protect\cite{Peeples2014} and \protect\cite{Bradford2015} are also presented. 
    Center panel: Specific star formation rate, sSFR($={\rm SFR}/M_{\rm *}$), versus stellar mass, $M_{\rm *}$, for the NIHAO (red circles) and NIHAO-LG simulations (purple and gold circles). 
    The underlying number density distribution of SFR and $M_{\rm *}$ measurements for MPA-JHU SDSS galaxies \protect{\citep{Kauffmann2003, Brinchmann2004}} are shown.
    Right-hand panel: MZR for the NIHAO (red circles) and NIHAO-LG simulations (purple and gold circles). Also presented is the observed relation from \protect\cite{Gallazzi2005}.}
    \label{fig:obs_sr}
\end{figure*}

We further tested the validity of NIHAO and NIHAO-LG simulations by comparing various observational scaling relations. 
The left-hand panel of \Fig{obs_sr} shows the cold gas ($T<20000\,{\rm K}$) content versus stellar mass in the NIHAO-LG simulations at redshift \rz within $0.2\,R_{\rm 200}$.
This choice of radius isolates the central parts of haloes as $\sim$90 per\,cent of the atomic and molecular gas resides within $0.2\,R_{\rm 200}$ of both the field galaxies and LG dwarfs.
Two observed cold gas mass-stellar mass relations from \cite{Peeples2014} and \cite{Bradford2015} are also presented.
The green squares represent the median cold gas mass-stellar mass relation in \cite{Peeples2014} using the observed data from \cite{McGaugh2005,McGaugh2012}, \cite{Leroy2008}, and \cite{Saintonge2011}.
The error bars represent the $16$per\,cent$-84$per\,cent percentile range of the observed data.
The cyan line and shaded region, taken from \cite{Bradford2015}, delineate the distribution of atomic gas mass for low mass galaxies ($\log M_* \leq 8.6$) in the Sloan Digital Sky Survey (SDSS; \citealt{sdssdr8}) and ALFALFA \citep{alfalfa}.
While the ALFALFA survey only detected \hi gas content, the total (atomic) gas content is calculated as $M_{\rm g} = 1.4M_{\hi}$ \citep{Oh2015}.

The high stellar mass galaxies (log$M_*\geq 8.5$) within the NIHAO and NIHAO-LG simulations agree well with observed cold gas content, while
the cold gas content in dwarfs from NIHAO and both NIHAO-LG simulations is higher than the observed relations of \cite{Peeples2014} and \cite{Bradford2015}.
The larger gas content in the simulated galaxies (field and LG) is due to the overcooling of gas.
However, given the large error bars (presenting the 16per\,cent$-84$per\,cent percentile range) at the low-mass end for \cite{Peeples2014}, a fraction of the simulated LG dwarfs agree with the observations.

We also compared the ${\rm sSFR}$--$M_{\rm *}$ relation in the central panel of \Fig{obs_sr} with observations.
The specific star formation rate, ${\rm sSFR} = {\rm SFR}/M_{\rm *}$, is the ratio of the average SFR within the last 100 Myr and the enclosed stellar mass within $0.2R_{\rm 200}$, $M_{\rm *}$.
In general, NIHAO and both NIHAO-LG simulations match the distribution of SDSS galaxies, albeit with simulated dwarfs falling in the outskirts of the observed distribution.
Given the uneven selection function of the simulated NIHAO systems and the different methods for measuring the observed quantities in \Fig{obs_sr}, comparing NIHAO galaxies with large-scale surveys is a non-trivial task.
However, even with this caveat, the overall agreement between NIHAO, NIHAO-LG and the SDSS is comforting.

The MZR relation of dwarf galaxies at redshift \rz can inform us about chemical evolution of the galaxies.
Most of the metals were formed within stars and are distributed into the galaxy via stellar feedback.
A comparison of the MZR between NIHAO and NIHAO-LG simulations and SDSS observations from \cite{Gallazzi2005} is presented in the right-hand panel of \Fig{obs_sr}.
While the observed data were only available for relatively massive systems ($\log(M_*/M_{\rm \odot}) \gtrsim 9.0$), both NIHAO and NIHAO-LG-simulated systems match observations well.
The MW and M31 analogues from both NIHAO-LG simulations also fall within the observed relation of \cite{Gallazzi2005}.
Small differences between simulations and observations slope measurements are expected from the sample sizes of our respective studies \citep{Sorce2016}.

A key feature of this study is to highlight the similarities and differences between the simulated NIHAO field systems and NIHAO-LG dwarfs.
Due to the near-identical dwarf galaxy properties in NIHAO-LG and NIHAO-LG(nmd) galaxies in Figs.\,\ref{fig:LG_pos}, \ref{fig:stellarhalo} and \ref{fig:obs_sr}; we only used dwarf galaxies from NIHAO-LG to compare with NIHAO field galaxies herein.
For general interest, the comparison of different properties between NIHAO field and NIHAO-LG(nmd) at redshift \rz is presented in \app{nmd}.

\section{Field and LG comparisons}
\label{sec:red0}

Below, we highlight similarities and differences between the NIHAO field and the NIHAO-LG dwarf samples. 
We begin with comparisons of the gas properties, specifically mass and metal content, followed by the metal content in stars of the simulated field and LG dwarfs.
Each comparison contains the calculation of the average difference between the NIHAO field and NIHAO-LG dwarfs. 
The average difference between the two NIHAO field and NIHAO-LG dwarfs is defined as:
\begin{equation}
    \langle\Delta (y_{\rm LG}|M_{\rm *})\rangle_{\rm LG} = \langle y_{\rm LG} - (\alpha_{\rm N}M_{\rm *,LG}+c_{\rm N})\rangle,
    \label{eq:avgres}
\end{equation}
\noindent where $y_{\rm LG}$ is a galaxy property (gas mass, average stellar metallicity, etc.), $\alpha_{\rm N}$ and $c_{\rm N}$ are the slope and intercept of the fitted scaling relation from NIHAO field sample, and $M_{\rm *,LG}$ is the stellar mass of the NIHAO-LG systems. 
The variables on the right-hand side of \Eq{avgres} all yield the average difference, $\langle\Delta (y_{\rm LG}|M_{\rm *})\rangle_{\rm LG}$.

\subsection{Gas mass}\label{sec:lg_gal} 

\begin{figure*}
    \centering
    \includegraphics[width=\linewidth]{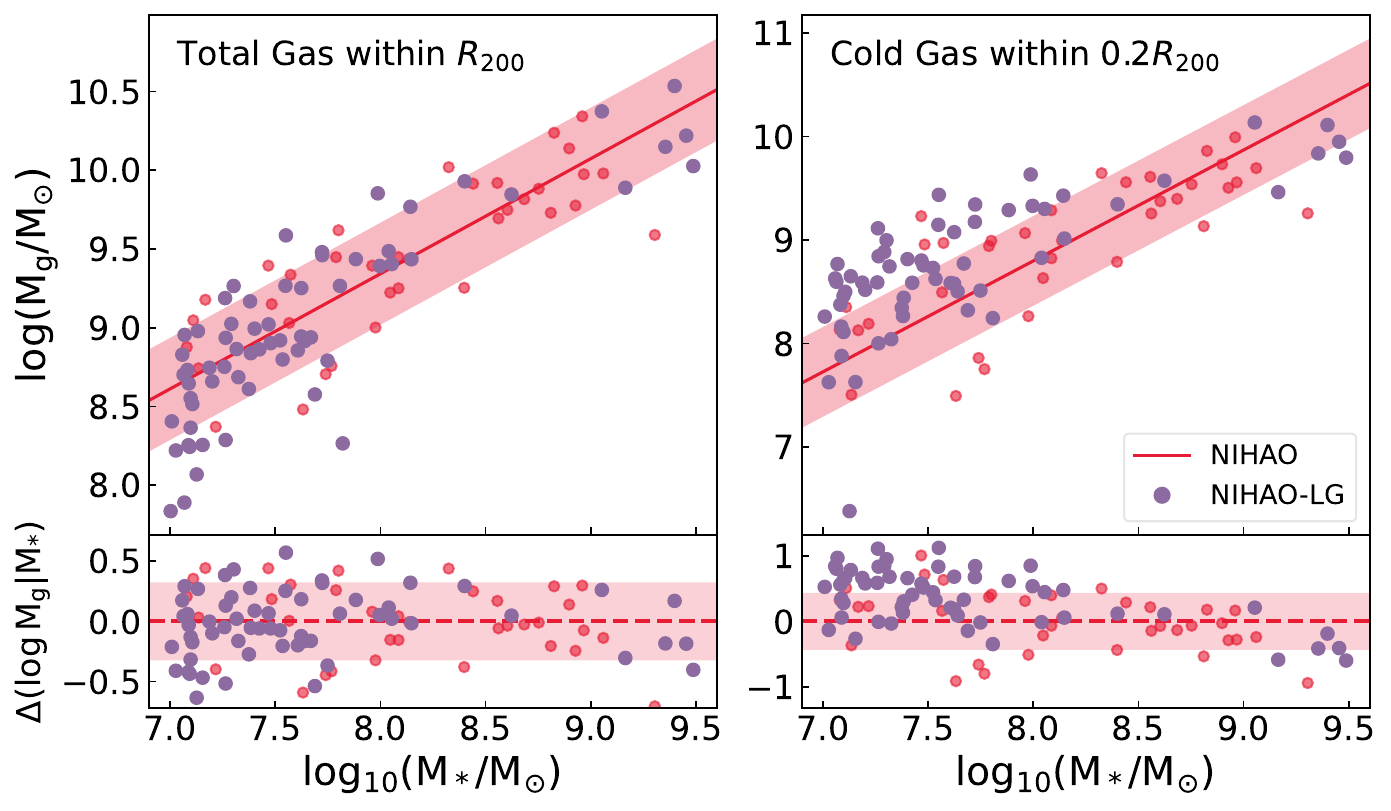}
    \caption{Total gas mass (left-hand panel) and cold gas mass  ($T<20000\,{\rm K}$; right-hand panel) versus stellar mass at \rz for NIHAO field (red circles) and NIHAO-LG (purple circles) dwarf galaxies. 
    The solid red line and shaded region represent a best fit of the NIHAO field dwarf galaxies and 1$\sigma$ scatter about that fit, respectively. 
    Total gas masses are measured within $\rm R_{\rm 200}$ while cold gas and stellar masses are measured within $\rm 0.2R_{\rm 200}$. 
    The residuals with respect to the NIHAO field dwarf best fit are shown in both bottom panels.}
    \label{fig:gasmass}
\end{figure*}

\begin{figure*}
    \centering
    \includegraphics[width=\linewidth]{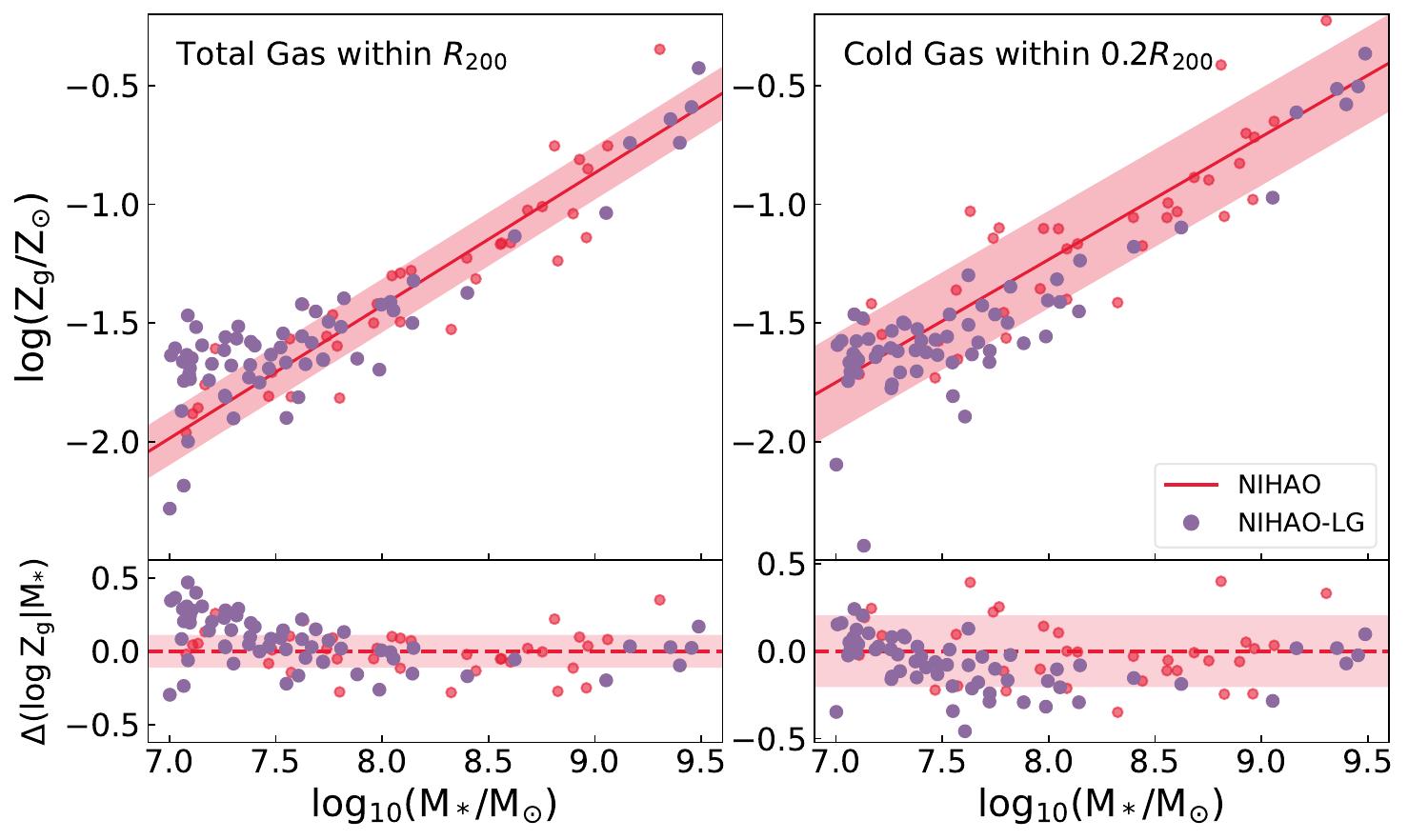}
    \caption{Average gas metallicity versus stellar mass for NIHAO field (red circles) and NIHAO-LG (purple circles) dwarf galaxies. The left- and right-hand panels show the metallicity of all the gas within $R_{\rm 200}$ and the metallicity of the cold gas within $0.2R_{\rm 200}$, respectively. 
    All other details are as in \Fig{gasmass}.}
    \label{fig:gasmetal}
\end{figure*}

We start with a comparison of the gas content of dwarf galaxies in the LG and in the field.
The left-hand panel of \Fig{gasmass} shows total gas mass within R$_{\rm 200}$ versus stellar mass (measured within $0.2R_{200}$) for the NIHAO field and NIHAO-LG dwarfs.
The red line and shaded region are linear best fit and scatter for the $M_{g}$--$M_{*}$ relation of the NIHAO field simulation.
It is found that the total gas content of the NIHAO-LG dwarfs lies within the $1\sigma$ scatter ($0.30\pm0.06\,$dex) of the NIHAO field galaxies.
A few LG dwarfs ($\log M_* \leq 7.8$) deviate from the observed trends for field dwarf systems which might be associated to random scatter about the NIHAO field galaxy relation.
Most of these NIHAO-LG dwarfs with lower gas content exhibit older, metal-poor stellar content as 50 per cent of the stellar mass was formed within the first $\sim$5\,Gyr of their formation.

The right-hand panel of \Fig{gasmass} shows the cold gas ($\rm T<20000\,K$) content in the field and LG dwarf systems at redshift \rz within $0.2\,R_{200}$.
The $1\sigma$ scatter for the NIHAO field dwarf relation is $\rm 0.46\pm0.10\, dex$, and the average difference with the NIHAO-LG dwarf population is calculated to be $\rm 0.41\pm 0.11$ (\Table{sim_sr}).
While the average difference for NIHAO-LG dwarfs, across the complete stellar mass range, is less than the scatter for the NIHAO field relation, LG dwarfs with $\log M_* \leq 8.0$ have a larger ($\sim$0.5\,dex) central cold gas content compared to field galaxies.
While the total gas distribution for field and LG are very similar; a larger fraction of the total gas within the NIHAO-LG dwarfs appears to be cold.
The residuals for the NIHAO-LG dwarfs in comparison to NIHAO field systems, shown in the bottom-right panel in \Fig{gasmass}, present a systematic trend for NIHAO-LG systems with an excess cold gas at the low-stellar mass end and a dearth of cold gas at the high-stellar mass end.

In summary, we have found similar distributions for the total gas contents within the NIHAO field and NIHAO-LG dwarfs. 
However, a significant fraction of the gas content within NIHAO-LG dwarfs exists as cold gas in the central parts. 
The unique gas content for the NIHAO-LG dwarfs should be connected to metal content and its evolution.
We explore the metal content of the NIHAO field and NIHAO-LG dwarfs next.

\subsection{Gas metallicity}

\Fig{gasmetal} shows a comparison of the gas metallicity between field NIHAO field and NIHAO-LG dwarf systems.
The left- and right-hand panels show the mean gas metallicity for all the gas within the halo and the mean gas metallicity for the cold gas within $0.2\,R_{\rm 200}$ respectively.
The field NIHAO systems have $Z_{\rm g}\propto M_{\rm *}^{0.56\pm0.05}$ with a scatter of $0.12\pm 0.03$ dex.
The NIHAO-LG dwarfs, on the other hand, exhibit a bimodal distribution in the $Z_{\rm g}$--$M_{\rm *}$.
Low stellar mass NIHAO-LG dwarfs ($\log(M_*/M_{\odot})\leq 8.0$) have a higher, though approximately constant, gas metallicity than NIHAO field with $\log(Z_g/Z_{\odot})\sim -1.6$.
High-stellar mass NIHAO-LG dwarfs follow the same trend and scatter as the NIHAO field galaxies.
The average difference between NIHAO field and NIHAO-LG systems of 0.18$\pm$0.06 for the $Z_{\rm g}$--$M_{\rm *}$ relation is larger than the 1$\sigma$ scatter for the NIHAO field relation.
This excess metallicity is addressed at greater length in \sec{insitu}.

The right-hand panel of \Fig{gasmetal} shows the cold gas metallicity versus stellar mass for NIHAO field and NIHAO-LG dwarfs.
The field sample follows a similar relation to total gas content in the halo, with $Z_{\rm g}\propto M_{\rm *}^{0.50\pm0.06}$ but with a larger scatter of $0.19\pm0.04$ dex. 
The NIHAO-LG dwarf galaxies fall within the field relation, presenting no difference in the central cold gas metallicity.
Indeed, the difference between NIHAO field and NIHAO-LG dwarfs is found to be negligible; equal to 0.00$\pm$0.04 (\Table{sim_sr}).
Comparing the left- and right-hand panels of \Fig{gasmetal}, we can conclude that metal-rich gas in the NIHAO-LG systems is found outside the central parts of dwarf halos, at radii greater than $0.2R_{\rm 200}$.
The source of the excess gas-phase metals could be indicative of recent inflowing metal-rich gas due to interactions in the LG which has yet to cool down, and/or the result of strong stellar feedback related outflows from the central regions of dwarf systems itself. 
To evaluate the dominant process, we compare field and LG dwarfs in the $M_{\rm g}$--$M_{*}$ and $Z_{\rm g}$--$M_{*}$ relations as a function of time in \sec{evol}.

\subsection{Stellar metallicities}

\begin{figure}
    \centering
    \includegraphics[width=\columnwidth]{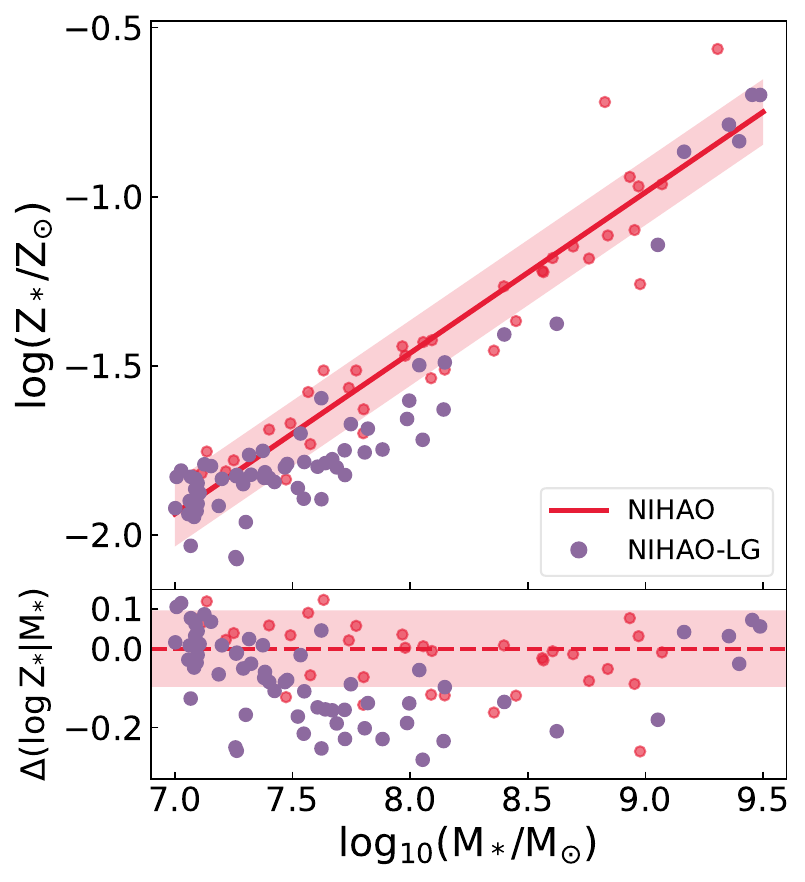}
    \caption{Average stellar metallicity versus stellar mass relation for the NIHAO field and NIHAO-LG dwarf galaxies. Both quantities were measured within a spherical radius of  $0.2R_{\rm 200}$. 
    All other details are as in \Fig{gasmass}.}
    \label{fig:starZ}
\end{figure}

\Fig{starZ} shows a comparison of stellar metallicity between NIHAO field and NIHAO-LG dwarf systems.
NIHAO field dwarfs follow $Z_{\rm *}\propto M_{\rm *}^{0.47\pm0.04}$ with a scatter of $\rm 0.09\pm0.02\, dex$.
NIHAO-LG dwarfs differ mildly ($\Delta Z_{\rm *}= 0.02\pm0.03$) from NIHAO field systems.
However, within the broad dispersions of each distributions, the field and NIHAO-LG dwarfs have statistically matching mean stellar metallicities. 
We also explored stellar velocity dispersion, mean stellar age, various measures of formation times and SFRs, and no striking 
differences between NIHAO field and NIHAO-LG dwarfs were found.
Within all practical measures, the NIHAO field and NIHAO-LG dwarfs have 
similar stellar populations.

The interested reader will find further comparisons between NIHAO field and NIHAO-LG dwarfs for the average stellar iron abundance in \app{iron}. 
Different methods for measuring iron abundance are also discussed.

\begin{table*}
\caption{Summary of scaling relations used for comparison of the three different simulated dwarf populations. Column (1) lists the scaling relation; columns (2–4) present the slope, zero-point, and scatter, respectively for the NIHAO field dwarfs; and column (5) shows the average difference between the NIHAO field and NIHAO-LG dwarf populations. If the value in column 5 is greater than the scatter of the NIHAO field scaling relations (column 4), the LG dwarf population is considered statistically different. The errors were bootstrapped over 2000 runs.}
\centering
\begin{tabular}{@{}lcccc@{}}
\toprule
                 & \multicolumn{3}{c}{NIHAO}                            & NIHAO-LG  \\ \midrule
Scaling Relation                 & Slope           & Zero--point      & Scatter         & Average Difference [dex] \\ 
(1)                              &(2)              &(3)               &(4)              & (5) \\ \midrule
$M_{\rm gas}-M_{*}$              & 0.73$\pm$0.09   & \phantom{-}3.47$\pm$0.73    & 0.30$\pm$0.06   & 0.00$\pm$0.05 \\
$M_{\rm cold gas}-M_{*}$         & 1.09$\pm$0.15   & \phantom{-}0.12$\pm$1.27    & 0.46$\pm$0.10   & 0.41$\pm$0.11 \\
$Z_{\rm gas}-M_{*}$              & 0.56$\pm$0.05   & -5.92$\pm$0.37   & 0.12$\pm$0.03   & 0.18$\pm$0.06 \\
$Z_{\rm cold gas}-M_{*}$         & 0.50$\pm$0.06   & -5.24$\pm$0.49   & 0.19$\pm$0.04   & 0.00$\pm$0.04 \\
$Z_{*}-M_{*}$                    & 0.47$\pm$0.04   & -5.25$\pm$0.31   & 0.09$\pm$0.02   & 0.02$\pm$0.03 \\ \bottomrule
\end{tabular}%
\label{tab:sim_sr}
\end{table*}

Finally for this section, \Table{sim_sr} presents a quantitative comparison of NIHAO field and NIHAO-LG dwarf populations.
For the scaling relations used here, the slope, zero-point, and scatter for the NIHAO field populations are shown.
The average difference between the NIHAO-LG and NIHAO field dwarf populations is also presented in \Table{sim_sr}.
That difference was calculated by randomly sampling the same number of NIHAO-LG dwarfs as the NIHAO field dwarfs (see \Table{sims}) and calculating the median residual with respect to the NIHAO field galaxy scaling relation as presented in \Eq{avgres}. 
This process was bootstrapped 2000 times to estimate the error of the average difference.
\Table{sim_sr} reiterates the larger differences in gas properties of the NIHAO-LG dwarfs relative to field systems, especially in the cold gas mass and the total gas metallicity.
The stellar properties, such as mean stellar metallicity, velocity dispersion, SFR, etc., of the NIHAO-LG dwarfs show little to no difference with NIHAO field dwarfs.

\section{Evolution of the LG}
\label{sec:evol}

We have found so far that the properties of simulated NIHAO field and NIHAO-LG dwarfs at low redshift (\rz) show non-negligible differences. 
NIHAO-LG dwarfs have larger (0.2 dex) stellar masses and a larger (0.41 dex) cold gas content within 0.2$R_{\rm 200}$ than field dwarfs at redshift . 
The hot gas content for NIHAO-LG dwarfs is also more metal-rich (0.18 dex) than in NIHAO field systems. 
In this section, we study the evolution of various galaxy properties in an attempt to isolate, if present, evolutionary differences between LG and field dwarf systems.

Motivated by our objective to study the similarities and differences between NIHAO field and NIHAO-LG dwarfs, we wish to trace the evolution of some key dwarf galaxy scaling relations with time.
In doing so, we must isolate specific times in the evolutionary history of NIHAO-LG galaxies where environment plays a key role.
Ultimately, we constrain the possible influence of LG environment on the evolution of NIHAO-LG dwarf galaxies.

The co-evolution of NIHAO-LG dwarfs in a LG-like environment should be apparent in the gas properties of the NIHAO-LG dwarfs relative to NIHAO field systems.
The full set of NIHAO-LG and NIHAO field dwarfs at all redshifts was compared using the average residual from the NIHAO field galaxy scaling relations.
The averages residual between NIHAO field and NIHAO-LG samples is calculated using \Eq{avgres} as a function of time.
The error on the average residuals are calculated as $\epsilon_{\rm LG}=\sigma(y_{\rm LG}|M_{\rm *})/\sqrt{N}$; where $\sigma(y_{\rm LG}|M_{\rm *})$ is the standard deviation of the residual from the field scaling relation and $N$ is the number of data points.
In this formalism, we have defined the LG and field dwarfs as different galaxy populations when $\epsilon_{\rm N} < \langle\Delta (y_{\rm LG}|M_{\rm *})\rangle_{\rm LG}(t)$, where $\epsilon_{\rm N}$ is the error of the linear fit to the NIHAO field galaxies.
In this comparison of the NIHAO-LG and NIHAO field dwarfs, the scatter of the field galaxy scaling relations is also presented. 
Due to their evolution with time, the slopes of nearly all scaling relations are also expected to evolve; therefore, the forward scatter ($\sigma_{\rm N, f}$) alone is not a robust comparison metric.
Instead, we use the orthogonal scatter, defined as $\sigma_{\rm N} = \sigma_{\rm N, f}/\sqrt{1+\alpha_{\rm N}^2}$, for our comparisons.

\subsection{Gas content}

\begin{figure}
    \centering
    \includegraphics[width=\columnwidth]{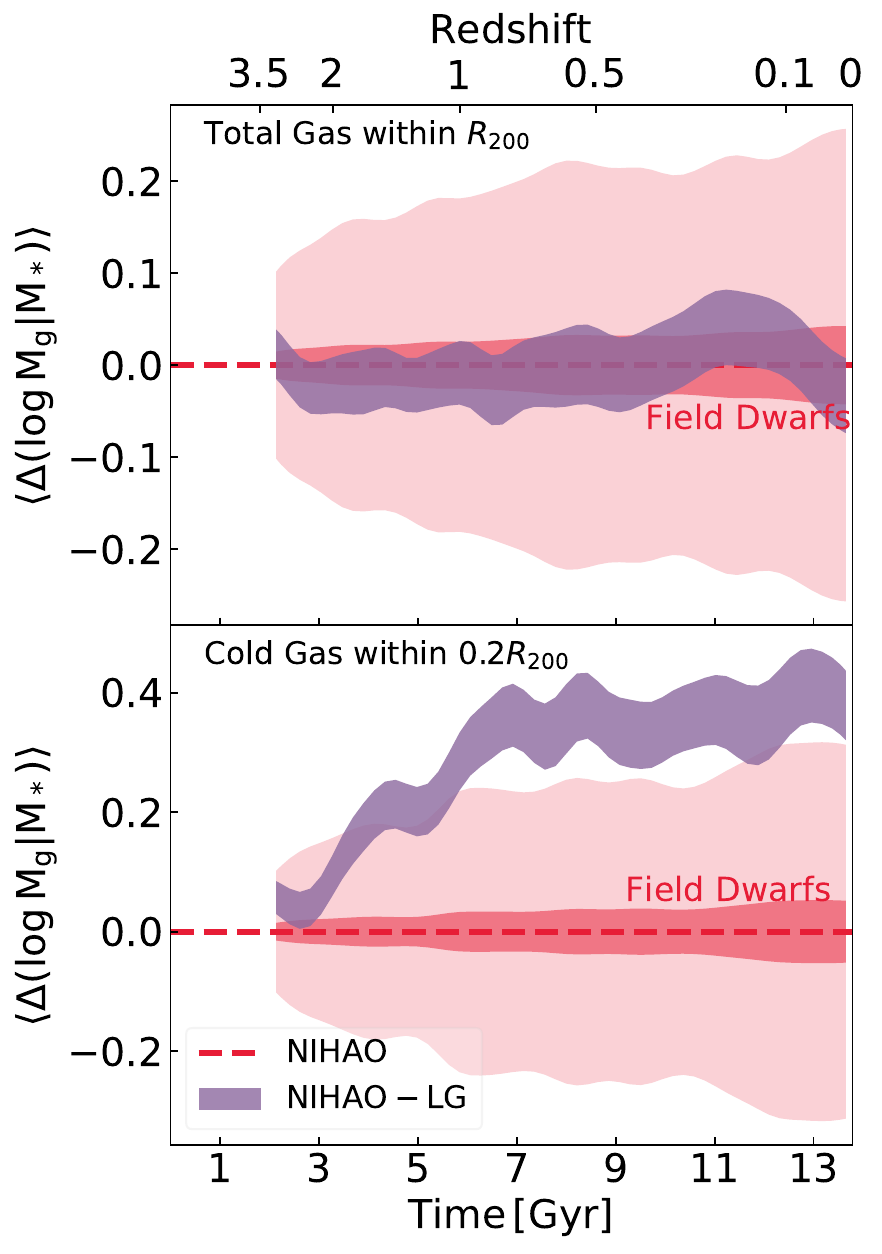}
    \caption{Evolution in the residuals of the gas mass–stellar mass relation as a function of time for NIHAO field (red) and NIHAO-LG (purple) dwarfs  $(7.0\leq \log(M_{\rm *}/M_{\rm \odot}) \leq 9.5)$ shown as shaded regions. 
    The gas mass-stellar mass relation of NIHAO field dwarfs is the horizontal line at $\langle \Delta (M_{\rm Gas}|M_*)\rangle = 0.0$, the dark red region gives the error on the average relation, and the lighter shaded region displays the $1\sigma$ scatter of the scaling relation as a function of time. 
    The top and bottom panels show the total and cold gas mass, respectively. 
    Redshift is shown on the top x-axis.}
    \label{fig:mgas_residuals}
\end{figure}

We study the evolution of gas properties, in particular mass and metallicity, for both NIHAO field and NIHAO-LG dwarfs. 
A comparison of the mass and/or metallicity of the total and/or gas in the NIHAO-LG haloes with the NIHAO field systems at multiple redshifts, can reveal the unique role of the Local Group environment in shaping properties of dwarf galaxies.
\Fig{mgas_residuals} shows the comparison of the field and LG systems as a function of time for the $M_{g}$--$M_{*}$ relation.
The top and bottom panels show all total gas content within $R_{\rm 200}$ and the cold gas content within $0.2R_{\rm 200}$, respectively.
The orthogonal scatter in the total $M_{g}$--$M_{*}$ relation within $R_{\rm 200}$ increases from $\sim$0.11 dex at redshift $z\sim 3.2$, to 0.26 dex at present day. 
Although not presented here, the slope of the relation shows little change over time.

Both NIHAO field and NIHAO-LG dwarfs have the same distribution of total gas content for the complete evolutionary history; the purple shaded band does indeed follow the dark red shaded region (see \Fig{mgas_residuals}).
The NIHAO field systems occupy the central regions of their respective DM halo and are fed gaseous material from the cosmic filaments, and infrequent gas-rich mergers.
Along with similar processes, NIHAO-LG dwarfs are expected to have modified gas content due to the high-density environment of the LG.
However, as depicted in \Fig{mgas_residuals}, the different environment of the LG does not play a significant role in altering the total gas mass of the NIHAO-LG dwarf galaxies \citep[see also][]{Sawala2012}.

While the evolution of the total gas content of the NIHAO-LG and NIHAO field systems is very similar, we find differences between the two dwarf populations for the central cold gas content.
For the NIHAO field dwarf sample, the slope and the scatter of the cold $M_{g}$--$M_{*}$ relation in the field evolve significantly with time (see bottom panel of \Fig{mgas_residuals}). 
The slope changes range from 0.80 at redshift $z\sim 3.2$ to 1.10 at present day, while the scatter grows from 0.09~dex at redshift $z\sim 3.2$ to 0.31~dex at redshift \rz.
Significant differences in the cold gas content between the NIHAO field and NIHAO-LG dwarfs emerge at redshift $z\lesssim 2$ and continue to grow until the present day. 
At redshift $z=0$, the NIHAO-LG dwarfs contain $\sim$0.4~dex more cold gas within $0.2R_{\rm 200}$ than the NIHAO field systems. 
The increasing cold gas mass in the NIHAO-LG dwarfs, compared to the field, is evidence of the unique role that the LG plays.
The excess central cold gas through the evolution of NIHAO-LG dwarfs is connected to the gas-phase metallicity evolution of the various star-formation episodes and interactions within the LG environment. 

\subsection{Chemical content}

\begin{figure}
    \centering
    \includegraphics[width=\columnwidth]{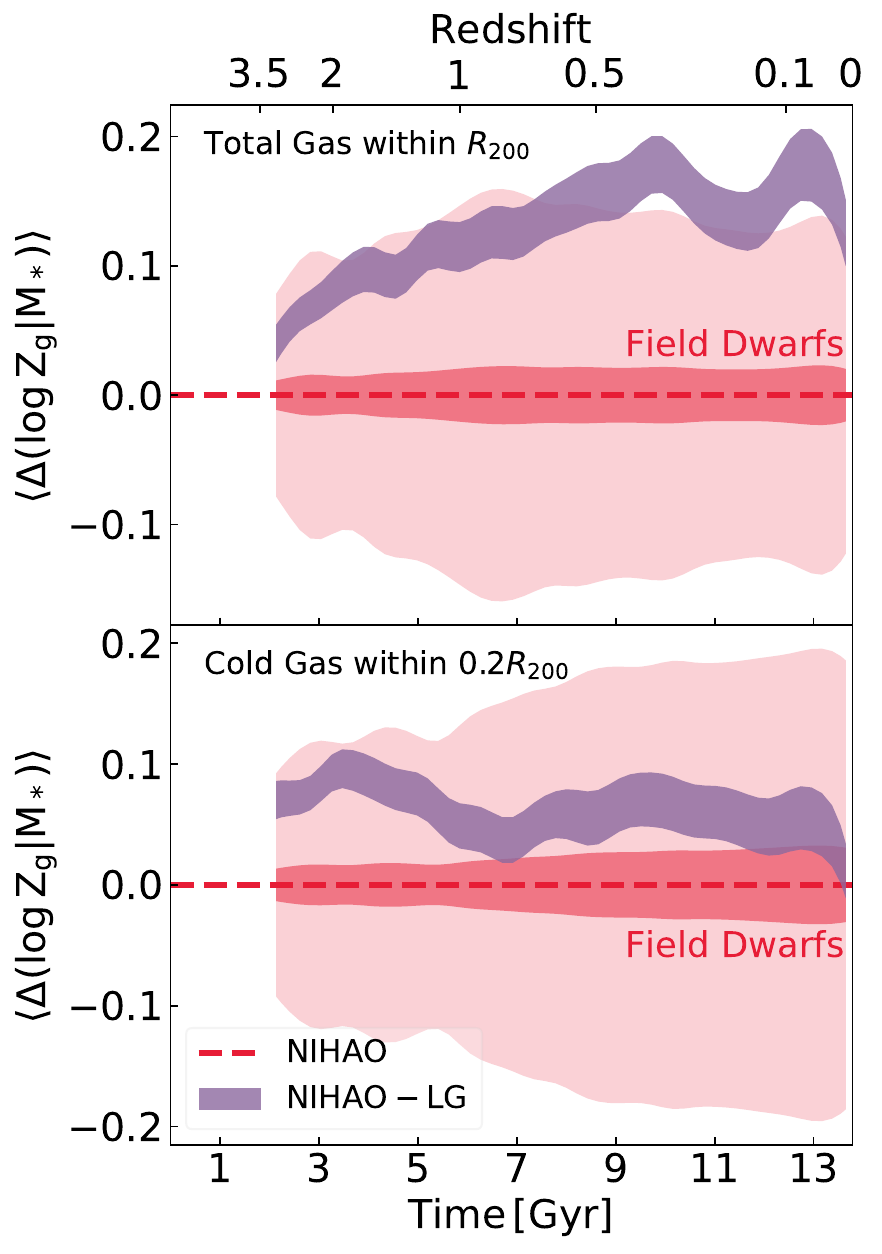}
    \caption{Same as \Fig{mgas_residuals} for gas metallicity.}
    \label{fig:zgas_residuals}
\end{figure}

We now consider the evolution of the gas-phase metal content in NIHAO field and NIHAO-LG dwarf galaxies. 
\Fig{zgas_residuals} shows the gas metallicity in the NIHAO field and NIHAO-LG dwarf samples over time.
The format of \Fig{zgas_residuals} is the same as \Fig{mgas_residuals}.
In the top panel, where gas metallicity of total gas content is computed within $R_{\rm 200}$, NIHAO field dwarfs display little variation in the scatter of the $Z_{\rm g}$--$M_{*}$ relation over time.
The NIHAO-LG simulations behave slightly differently with time showing dwarfs with more metal-rich gas in the halo than the NIHAO field systems.

From redshift $z\sim 1$ to present day, the \nlg dwarfs possess a more metal rich gas content (by 0.15~dex, or $3\,\sigma$) than the NIHAO field systems.
While the total gas mass of the NIHAO field and NIHAO-LG dwarfs evolve in similar ways (see top panel of \Fig{mgas_residuals}), the gas content of the NIHAO-LG dwarfs is significantly more metal-enriched.
The excess gas-phase metals in NIHAO-LG dwarfs can be attributed to either the interactions within the high-density environment of the LG and/or due to star-formation driven feedback.
The next section will investigate the dominance of the environmental (accretion in a high-density environment) and \textit{in-situ} (star formation-driven) processes for the metallicity evolution of the LG.

The bottom panel in \Fig{zgas_residuals} depicts the cold gas metallicity evolution of the NIHAO field and NIHAO-LG dwarf galaxies.
Unlike the $Z_{\rm g}$--$M_{*}$ relation for total gas, the cold gas $Z_{\rm g}$--$M_{*}$ relation for the NIHAO field shows increasing scatter as a function of time. 
Once again, the NIHAO-LG dwarfs behave differently; over their complete evolution history, NIHAO-LG dwarfs have more cold gas metallicity relative to the NIHAO field systems. 
The excess metal in the gas within the halo translates into metal-rich cold gas in the central parts ($0.2R_{\rm 200}$).
The difference between the NIHAO field and NIHAO-LG dwarfs for cold gas metallicity decreases over redshift $z\sim 1-2$.
During this period, the total gas metallicity (top panel of \Fig{zgas_residuals}) difference between NIHAO field and NIHAO-LG dwarfs continues to grow.

For redshift $z<1$, the difference between NIHAO field and NIHAO-LG for cold gas metallicity remains approximately constant. 
\nlg dwarfs show marginal evolutionary differences from the NIHAO field dwarfs for stellar properties, especially for the average stellar metallicity.
The lack of evolution in the cold gas metallicity and similar stellar properties is linked to the lack of late star-formation activity in NIHAO-LG dwarfs. 
Star-formation quenching hinders the recycling of metals in gas and yields a static metal content in the NIHAO-LG cold gas content.

\subsection{Metal enrichment: \textit{in-situ} versus environment}
\label{sec:insitu}

\begin{figure*}
    \centering
    \includegraphics[width=\linewidth]{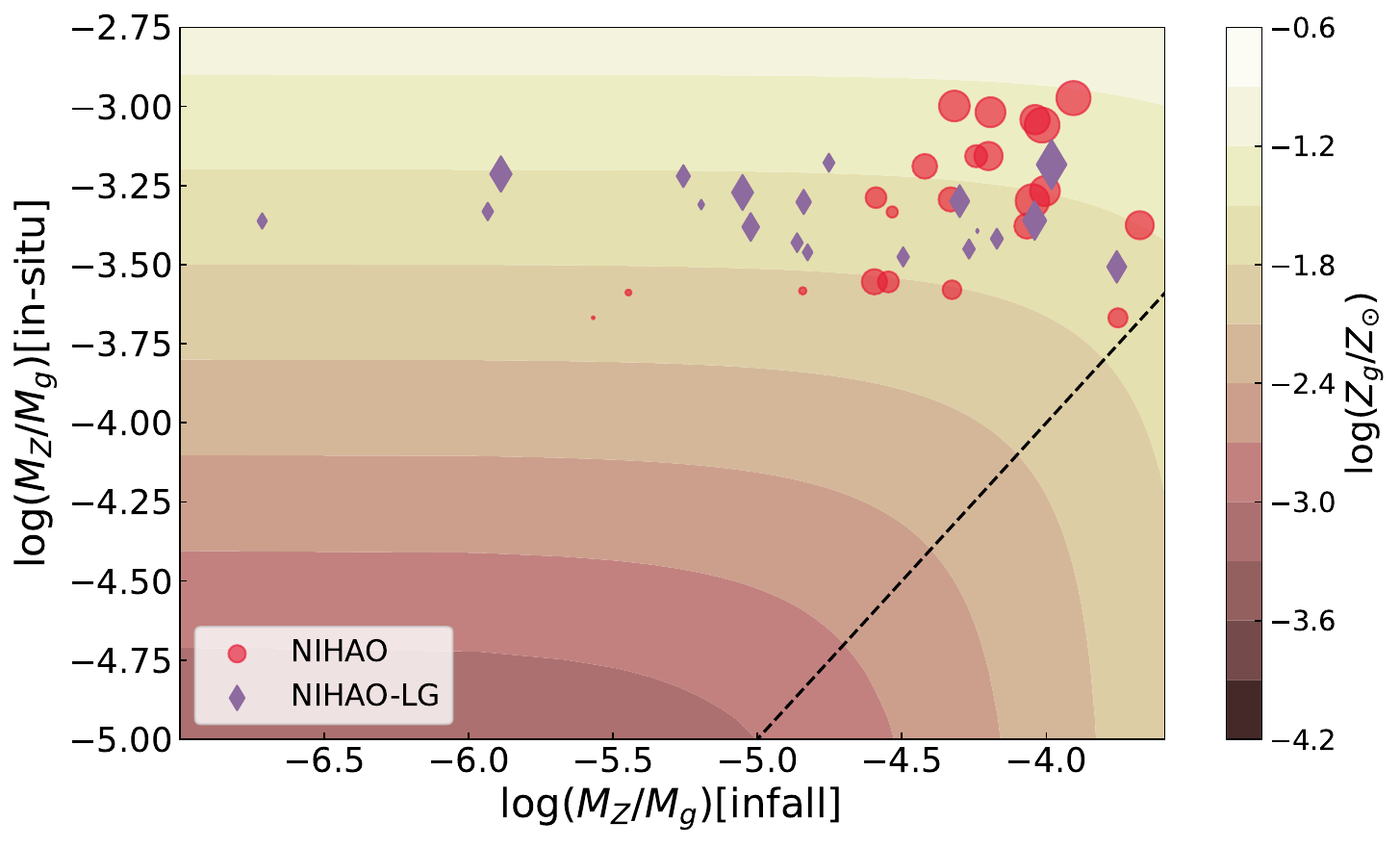}
    \caption{Amount of gas-phase metal mass accreted from infall versus gas-phase metal mass generated \textit{in-situ}, for simulated dwarf haloes normalized by the total gas mass at \rz. 
    Red circles correspond to NIHAO field systems, while purple diamonds represent the NIHAO-LG dwarfs. 
    The size of the each point corresponds to the stellar mass at \rz. 
    The contours display present-day constant gas phase metallicity (see color scale to the right). 
    The dashed line represents the line of 50\% \textit{in-situ} metal evolution.
    Only NIHAO-LG galaxies found 1.5$\sigma$ outside of the NIHAO field $Z_{\rm gas}-M_{\rm *}$ relation are shown (see \Fig{gasmetal}).}
    \label{fig:insitu_infall}
\end{figure*}

The metal evolution for these simulated dwarf haloes can come from two sources: \textit{in-situ} star formation (and its associated feedback) as well as interactions within a dense environment of the LG.
\Fig{insitu_infall}, which attempts to isolate which of these processes is dominant, shows the comparison of gas-phase metals accreted onto a simulated dwarf versus the gas-phase metals produced in-situ, normalized by the gas mass at redshift \rz.
It should be noted that only NIHAO-LG dwarfs lying 1.5$\sigma$ outside of the NIHAO field $Z_{\rm gas}-M_{\rm *}$ relation presented in \Fig{gasmetal} are included in \Fig{insitu_infall}.

To calculate the accreted and \textit{in-situ} metal masses of a dwarf halo, all gas particles within a halo (at redshift \rz) are traced back to the time when they became bound to the halo progenitor.
The metal fraction for gas particles at infall translates to the accreted metal mass and the metals created \textit{in-situ}, ($M_{Z}(\rm{in-situ})$), according to:
\begin{equation}
    M_{Z}({\rm in-situ}) = M_{Z,0} - M_{Z}(@~{\rm infall}),
    \label{eq:insitu_metal}
\end{equation}
where $M_{Z,0}$ is the present-day metal mass for gas particles.
\Fig{insitu_infall} also shows contours of constant gas-phase metallicity, $\log(Z_g/Z_{\odot})$, at redshift \rz calculated as:
\begin{equation}
\begin{split}
    \log\Bigg(\frac{Z_g}{Z_{\odot}}\Bigg) &= 
    \log\Bigg(\frac{M_{Z, 0}}{M_g}\frac{1}{Z_{\odot}}\Bigg)\\
    & = \log\Bigg(\Bigg[\frac{M_{Z}(@~{\rm infall})}{M_g}+\frac{M_{Z}({\rm in-situ})}{M_g}\Bigg]\frac{1}{Z_{\odot}}\Bigg).
\end{split}
\label{eq:cons_gasz}
\end{equation}
Horizontal or vertical displacements (i.e., metallicity evolution) of dwarf haloes along the contours in \Fig{insitu_infall} are predominantly driven by accretion and/or stellar feedback related outflows, respectively.
Galaxies located on the curved part of these contours also have comparable contributions to their metallicity evolution from both accretion and \textit{in-situ} processes.
It should be noted that all galaxies (NIHAO and NIHAO-LG) produce more gas-phase metals through star-formation processes (\textit{in-situ} production) than accretion.
The dashed line represents 50 per\,cent of the metals created through \textit{in-situ} processes. 
All galaxies (NIHAO field and NIHAO-LG) lie on the left-hand side of the dashed line.

We find that nearly all NIHAO field dwarfs lie on the curved region of the constant gas-phase metallicity contours showing contributions from both \textit{in-situ} and accretion processes. 
NIHAO haloes with larger stellar mass reside on contours of larger metallicity (a different way to represent the field $Z_{\rm gas}-M_{\rm *}$ relation), while NIHAO-LG dwarf haloes lie on the horizontal parts of constant contours ($\log(Z_g/Z_{\odot}\sim -1.7)$).
We find that all  NIHAO-LG dwarfs (which are outliers of the field $Z_{\rm gas}-M_{\rm *}$ relation) have constant amounts of metals created via in-situ processes but varying contributions from the environment.

Analysing the accretion of gas-phase metals as a function of time, we find that NIHAO-LG dwarfs accrete gas with high metal content at late times ($t\sim7\,{\rm Gyr}$).
We have also traced the origins of gas particles with high metal fractions.
We find that only three NIHAO-LG dwarf haloes with $\log(M_{*}/M_{\odot}<8.5)$ have gas particles that evolved through star formation in the MW/M31 analogues.
With the MW/M31 analogues playing minimal roles in the metallicity evolution, the role of environment in this case is a result of the co-evolution of NIHAO-LG dwarfs in a high-density region in the Universe. 
Throughout their evolution, LG dwarfs co-evolve whilst sharing high metallicity gas released through active star formation feedback. 
Given that the accretion in LG dwarfs occurs at a late time, the gas has not had time to cool (the cooling time typically exceeds the Hubble time) and participate in star formation activities.

As a result, no differences are found for the metal content of stars between field and LG dwarfs (see \Fig{starZ}).
Finally, some NIHAO-LG dwarf halos, show negligible infall metals but high \textit{in-situ} gas-phase metals.
Such galaxies accrete a large number of gas particles with low metal fractions early in their evolution and process gas through star formation, resulting in higher metallicities.

We also found that NIHAO-LG dwarf with excess gas-phase metals are randomly distributed through the simulated LG region; no correlations exist between the excess gas metallicity and distances from MW/M31 analogues or the barycenter of the LG. 
This is further evidence that MW/M31 analogues play negligible roles in the metallicity evolution of NIHAO-LG dwarfs. 
It is the full dwarf population in simulated LG regions which goes through star formation at early times to process gas and then exchange it with other NIHAO-LG dwarfs at late times. 
This creates a unique gas property for NIHAO-LG dwarf populations with respect to NIHAO field systems.

\section{Summary and Conclusions}
\label{sec:conclusion}

The Local Group (LG) is a superb laboratory for the study of the formation and evolution of the dwarf galaxies; the most abundant cosmological structures in the Universe. 
In this paper, we have examined whether the LG provides an unbiased foil for studies of galaxy formation and evolution.
In this comparative study, we have highlighted the similarities and differences between simulated LG and field dwarf galaxies as a function of time/redshift and compared with observations. 
Our field control sample relied on dwarf haloes from the NIHAO high resolution zoom-in simulations. 
For the LG dwarfs, we used the constrained LG simulations run with NIHAO hydrodynamics.

Present-day NIHAO field and NIHAO-LG dwarfs were found to have similar stellar populations; properties like stellar velocity dispersion, mean stellar age, accretion times, and SFRs, etc.
Relative to field systems, NIHAO-LG dwarfs show small evolution in cold gas metallicity with time. 
This is understood in the following way: While the high metallicity gas in the halo should allow for gas to cool efficiently and lead to star-formation events, stellar feedback and outflows in the dwarf systems suppress any further stellar evolution leading to a quenched system and locking metals in the hot halo gas.
Therefore, the stellar and cold gas evolution of NIHAO-LG dwarfs is strongly dictated by \textit{in-situ} processes such as stellar feedback, winds, and photoevaporation due to re-ionization, which are not affected by environment \citep{Sawala2012}.
A significant fraction of the dwarfs in our constrained LG simulations build up their stellar content early in their evolution history and remain quenched thereafter \citep[eg.][]{Sand2010, Simon2021}.

At all times, NIHAO-LG and NIHAO field dwarfs have very similar total gas content. 
In other words, the total gas content of the LG dwarfs is not influenced by environment, in agreement with findings of \cite{Sawala2012}.
However, a larger fraction of the gas within NIHAO-LG dwarfs is cold and resides within $0.2R_{\rm 200}$.
The larger cold gas content is correlated to the higher metal content (relative to the NIHAO field dwarfs) of the gas in the halo; which cools down and enhances the metal content at early times. 
The higher gas metal content of the NIHAO-LG dwarfs is expected to be accumulated via interactions within the LG which allows the NIHAO-LG dwarfs to exchange the heavy metal generated during star-formation episodes in other LG systems. 
The time evolution of the cold gas mass within NIHAO-LG dwarfs shows departures from the field sample around redshift $z\sim1-2$.
This excess cold gas content in NIHAO-LG dwarf through time is related to the metal evolution of the hot gas in NIHAO-LG dwarfs as well; departure for NIHAO-LG gas metallicity emerges around redshift  $z\sim1-2$.
Indeed, relative to NIHAO field systems, NIHAO-LG dwarf possess excess metals in the hot gas phase at $R>0.2R_{\rm 200}$ at present day.

We isolate the dominance of \textit{in-situ} metal evolution and/or impact of the LG environment. 
Most NIHAO-LG dwarf galaxies present excess gas-phase metals relative to simulated NIHAO field dwarfs; these metal-rich gas reside at radii $>0.2\,R_{
\rm 200}$.
While environment does play a role, the direct influence or presence of the massive haloes (MW/M31) is found to be insignificant; rather, the interaction of dwarfs in a high-density environment such as the LG is the dominant factor. 
We found that these NIHAO-LG dwarfs accrete high-metallicity gas, processed in other NIHAO-LG dwarfs, at late times ($t\sim 7{\rm Gyr}$).
Because of the late accretion times, the NIHAO-LG dwarf hot gas-phase metals have not had time to cool and participate in star-formation activity.
As a result, the gas metallicity evolution of the LG is not reflected in stellar properties of NIHAO-LG dwarfs.
The NIHAO-LG dwarfs with excess gas-phase metals are found to have stochastic distribution through the simulated regions; that is, no correlations exist between the excess gas metallicity and the central parts of the simulation LG region.

Given the similar stellar properties of simulated dwarfs, high-resolution dwarf simulations such as NIHAO \citep{nihao_main} and FIRE \citep{Garrison2019} may then be compared with the general dwarf population.
Our results have demonstrated the value of LG observations, specifically their stellar properties, as constraints for the overall dwarf populations in our Universe.
The unique aspects seen in the gas properties of the LG can be examined using high-resolution constrained LG simulations \citep{CLUES2010, Sorce2016b, Carlesi2016, Libeskind2020}.
Furthermore, with the advent of next generation telescopes such as JWST, Euclid, Rubin/LSST, and SKA, observational signatures of pre-enrichment can ultimately be teased out of high redshift LG analogs and their accompanying dwarfs.

\section*{Data Availability}
The data underlying this article will be shared upon request to the corresponding author(s).

\section*{Acknowledgements}

NA, SC, and CS are grateful to the Natural Sciences and Engineering Research Council of Canada, the Ontario Government, and Queen's University for generous support through various scholarships and grants.
NA also acknowledges support from the Arthur B. McDonald Canadian Astroparticle Physics Research Institute and the Canada First Research Excellence Fund.
TB acknowledges support from the European Research Council under ERC-CoG grant CRAGSMAN-646955. 
NIL acknowledges financial support from the Project IDEXLYON at the University of Lyon under the Investments for the Future Program (ANR-16-IDEX-0005).
JGS acknowledges support from the ANR LOCALIZATION project, grant ANR-21-CE31-0019 of the French Agence Nationale de la Recherche.
GY thanks the Spanish Ministry of  Science, Innovation and Universities  (MICIU/FEDER)  for financial support under research grant PGC2018-094975-C2.
YH has been partially supported by the Israel Science Foundation grant ISF 1358/18.
Barbara Catinella, Luca Cortese, Ananthan Karunakaran, and Kristine Spekkens provided valuable references and discussion about our observational comparisons to simulated data. 
Aaron Dutton is also thanked for earlier contributions and comments relevant to this work.
Our conscientious referee also offered most valuable comments and suggestions which improved the presentation and content of this paper.
The research was performed using the {\scriptsize PYNBODY} package \citep{Pontzen2013}, S{\scriptsize CI}P{\scriptsize Y} \citep{scipy}, and N{\scriptsize UM}P{\scriptsize Y} \citep{numpy} and used {\scriptsize MATPLOTLIB} \citep{matplotlib} for all graphical representation.
The authors gratefully acknowledge the Gauss Centre for Supercomputing e.V. (www.gauss-centre.eu) for funding this project by providing computing time on the GCS Supercomputer SuperMUC at Leibniz Supercomputing Centre (www.lrz.de).
This research was carried out on the High Performance Computing resources at New York University Abu Dhabi. 
We greatly appreciate the contributions of all these computing allocations.


\bibliographystyle{mnras}
\bibliography{references} 

\begin{thebibliography}{}
\makeatletter
\relax
\def\mn@urlcharsother{\let\do\@makeother \do\$\do\&\do\#\do\^\do\_\do\%\do\~}
\def\mn@doi{\begingroup\mn@urlcharsother \@ifnextchar [ {\mn@doi@}
  {\mn@doi@[]}}
\def\mn@doi@[#1]#2{\def\@tempa{#1}\ifx\@tempa\@empty \href
  {http://dx.doi.org/#2} {doi:#2}\else \href {http://dx.doi.org/#2} {#1}\fi
  \endgroup}
\def\mn@eprint#1#2{\mn@eprint@#1:#2::\@nil}
\def\mn@eprint@arXiv#1{\href {http://arxiv.org/abs/#1} {{\tt arXiv:#1}}}
\def\mn@eprint@dblp#1{\href {http://dblp.uni-trier.de/rec/bibtex/#1.xml}
  {dblp:#1}}
\def\mn@eprint@#1:#2:#3:#4\@nil{\def\@tempa {#1}\def\@tempb {#2}\def\@tempc
  {#3}\ifx \@tempc \@empty \let \@tempc \@tempb \let \@tempb \@tempa \fi \ifx
  \@tempb \@empty \def\@tempb {arXiv}\fi \@ifundefined
  {mn@eprint@\@tempb}{\@tempb:\@tempc}{\expandafter \expandafter \csname
  mn@eprint@\@tempb\endcsname \expandafter{\@tempc}}}

\bibitem[\protect\citeauthoryear{{Aihara} et~al.,}{{Aihara}
  et~al.}{2011}]{sdssdr8}
{Aihara} H.,  et~al., 2011, \mn@doi [\apjs] {10.1088/0067-0049/193/2/29}, \href
  {https://ui.adsabs.harvard.edu/abs/2011ApJS..193...29A} {193, 29}

\bibitem[\protect\citeauthoryear{{Behroozi}, {Wechsler}  \&
  {Conroy}}{{Behroozi} et~al.}{2013}]{Behroozi2013}
{Behroozi} P.~S.,  {Wechsler} R.~H.,   {Conroy} C.,  2013, \mn@doi [\apj]
  {10.1088/0004-637X/770/1/57}, \href
  {https://ui.adsabs.harvard.edu/abs/2013ApJ...770...57B} {770, 57}

\bibitem[\protect\citeauthoryear{{Ben{\'\i}tez-Llambay}, {Navarro}, {Abadi},
  {Gottl{\"o}ber}, {Yepes}, {Hoffman}  \& {Steinmetz}}{{Ben{\'\i}tez-Llambay}
  et~al.}{2015}]{benitez2015}
{Ben{\'\i}tez-Llambay} A.,  {Navarro} J.~F.,  {Abadi} M.~G.,  {Gottl{\"o}ber}
  S.,  {Yepes} G.,  {Hoffman} Y.,   {Steinmetz} M.,  2015, \mn@doi [\mnras]
  {10.1093/mnras/stv925}, \href
  {https://ui.adsabs.harvard.edu/abs/2015MNRAS.450.4207B} {450, 4207}

\bibitem[\protect\citeauthoryear{{Ben{\'\i}tez-Llambay}, {Navarro}, {Abadi},
  {Gottl{\"o}ber}, {Yepes}, {Hoffman}  \& {Steinmetz}}{{Ben{\'\i}tez-Llambay}
  et~al.}{2016}]{benitez2016}
{Ben{\'\i}tez-Llambay} A.,  {Navarro} J.~F.,  {Abadi} M.~G.,  {Gottl{\"o}ber}
  S.,  {Yepes} G.,  {Hoffman} Y.,   {Steinmetz} M.,  2016, \mn@doi [\mnras]
  {10.1093/mnras/stv2722}, \href
  {https://ui.adsabs.harvard.edu/abs/2016MNRAS.456.1185B} {456, 1185}

\bibitem[\protect\citeauthoryear{{Bland-Hawthorn} \&
  {Gerhard}}{{Bland-Hawthorn} \& {Gerhard}}{2016}]{Bland-Hawthorn2016}
{Bland-Hawthorn} J.,  {Gerhard} O.,  2016, \mn@doi [\araa]
  {10.1146/annurev-astro-081915-023441}, \href
  {https://ui.adsabs.harvard.edu/abs/2016ARA&A..54..529B} {54, 529}

\bibitem[\protect\citeauthoryear{{Blank}, {Meier}, {Macci{\`o}}, {Dutton},
  {Dixon}, {Soliman}  \& {Kang}}{{Blank} et~al.}{2021}]{Blank2021}
{Blank} M.,  {Meier} L.~E.,  {Macci{\`o}} A.~V.,  {Dutton} A.~A.,  {Dixon}
  K.~L.,  {Soliman} N.~H.,   {Kang} X.,  2021, \mn@doi [\mnras]
  {10.1093/mnras/staa2670}, \href
  {https://ui.adsabs.harvard.edu/abs/2021MNRAS.500.1414B} {500, 1414}

\bibitem[\protect\citeauthoryear{{Blanton} \& {Moustakas}}{{Blanton} \&
  {Moustakas}}{2009}]{Blanton2009}
{Blanton} M.~R.,  {Moustakas} J.,  2009, \mn@doi [\araa]
  {10.1146/annurev-astro-082708-101734}, \href
  {https://ui.adsabs.harvard.edu/abs/2009ARA&A..47..159B} {47, 159}

\bibitem[\protect\citeauthoryear{{Bradford}, {Geha}  \& {Blanton}}{{Bradford}
  et~al.}{2015}]{Bradford2015}
{Bradford} J.~D.,  {Geha} M.~C.,   {Blanton} M.~R.,  2015, \mn@doi [\apj]
  {10.1088/0004-637X/809/2/146}, \href
  {https://ui.adsabs.harvard.edu/abs/2015ApJ...809..146B} {809, 146}

\bibitem[\protect\citeauthoryear{{Brinchmann}, {Charlot}, {White}, {Tremonti},
  {Kauffmann}, {Heckman}  \& {Brinkmann}}{{Brinchmann}
  et~al.}{2004}]{Brinchmann2004}
{Brinchmann} J.,  {Charlot} S.,  {White} S.~D.~M.,  {Tremonti} C.,  {Kauffmann}
  G.,  {Heckman} T.,   {Brinkmann} J.,  2004, \mn@doi [\mnras]
  {10.1111/j.1365-2966.2004.07881.x}, \href
  {https://ui.adsabs.harvard.edu/abs/2004MNRAS.351.1151B} {351, 1151}

\bibitem[\protect\citeauthoryear{{Brown} et~al.,}{{Brown}
  et~al.}{2012}]{Brown2012}
{Brown} T.~M.,  et~al., 2012, \mn@doi [\apjl] {10.1088/2041-8205/753/1/L21},
  \href {https://ui.adsabs.harvard.edu/abs/2012ApJ...753L..21B} {753, L21}

\bibitem[\protect\citeauthoryear{{Buck}}{{Buck}}{2020}]{Buck2020a}
{Buck} T.,  2020, \mn@doi [\mnras] {10.1093/mnras/stz3289}, \href
  {https://ui.adsabs.harvard.edu/abs/2020MNRAS.491.5435B} {491, 5435}

\bibitem[\protect\citeauthoryear{{Buck}, {Macci{\`o}}, {Obreja}, {Dutton},
  {Dom{\'\i}nguez-Tenreiro}  \& {Granato}}{{Buck} et~al.}{2017}]{Buck2017}
{Buck} T.,  {Macci{\`o}} A.~V.,  {Obreja} A.,  {Dutton} A.~A.,
  {Dom{\'\i}nguez-Tenreiro} R.,   {Granato} G.~L.,  2017, \mn@doi [\mnras]
  {10.1093/mnras/stx685}, \href
  {https://ui.adsabs.harvard.edu/abs/2017MNRAS.468.3628B} {468, 3628}

\bibitem[\protect\citeauthoryear{{Buck}, {Macci{\`o}}, {Dutton}, {Obreja}  \&
  {Frings}}{{Buck} et~al.}{2019}]{Buck2019}
{Buck} T.,  {Macci{\`o}} A.~V.,  {Dutton} A.~A.,  {Obreja} A.,   {Frings} J.,
  2019, \mn@doi [\mnras] {10.1093/mnras/sty2913}, \href
  {https://ui.adsabs.harvard.edu/abs/2019MNRAS.483.1314B} {483, 1314}

\bibitem[\protect\citeauthoryear{{Buck}, {Obreja}, {Macci{\`o}}, {Minchev},
  {Dutton}  \& {Ostriker}}{{Buck} et~al.}{2020}]{Buck2020b}
{Buck} T.,  {Obreja} A.,  {Macci{\`o}} A.~V.,  {Minchev} I.,  {Dutton} A.~A.,
  {Ostriker} J.~P.,  2020, \mn@doi [\mnras] {10.1093/mnras/stz3241}, \href
  {https://ui.adsabs.harvard.edu/abs/2020MNRAS.491.3461B} {491, 3461}

\bibitem[\protect\citeauthoryear{{Buck}, {Rybizki}, {Buder}, {Obreja},
  {Macci{\`o}}, {Pfrommer}, {Steinmetz}  \& {Ness}}{{Buck}
  et~al.}{2021}]{Buck2021}
{Buck} T.,  {Rybizki} J.,  {Buder} S.,  {Obreja} A.,  {Macci{\`o}} A.~V.,
  {Pfrommer} C.,  {Steinmetz} M.,   {Ness} M.,  2021, arXiv e-prints, \href
  {https://ui.adsabs.harvard.edu/abs/2021arXiv210303884B} {p. arXiv:2103.03884}

\bibitem[\protect\citeauthoryear{{Carlesi} et~al.,}{{Carlesi}
  et~al.}{2016}]{Carlesi2016}
{Carlesi} E.,  et~al., 2016, \mn@doi [\mnras] {10.1093/mnras/stw357}, \href
  {https://ui.adsabs.harvard.edu/abs/2016MNRAS.458..900C} {458, 900}

\bibitem[\protect\citeauthoryear{{Chabrier}}{{Chabrier}}{2003}]{Chabrier2003}
{Chabrier} G.,  2003, \mn@doi [\pasp] {10.1086/376392}, \href
  {https://ui.adsabs.harvard.edu/abs/2003PASP..115..763C} {115, 763}

\bibitem[\protect\citeauthoryear{{Clemens}, {Bressan}, {Nikolic}, {Alexander},
  {Annibali}  \& {Rampazzo}}{{Clemens} et~al.}{2006}]{Clemens2006}
{Clemens} M.~S.,  {Bressan} A.,  {Nikolic} B.,  {Alexander} P.,  {Annibali} F.,
    {Rampazzo} R.,  2006, \mn@doi [\mnras] {10.1111/j.1365-2966.2006.10530.x},
  \href {https://ui.adsabs.harvard.edu/abs/2006MNRAS.370..702C} {370, 702}

\bibitem[\protect\citeauthoryear{{Cluver} et~al.,}{{Cluver}
  et~al.}{2020}]{Cluver2020}
{Cluver} M.~E.,  et~al., 2020, \mn@doi [\apj] {10.3847/1538-4357/ab9cb8}, \href
  {https://ui.adsabs.harvard.edu/abs/2020ApJ...898...20C} {898, 20}

\bibitem[\protect\citeauthoryear{{Danieli}, {van Dokkum}  \&
  {Conroy}}{{Danieli} et~al.}{2018}]{Danieli2018}
{Danieli} S.,  {van Dokkum} P.,   {Conroy} C.,  2018, \mn@doi [\apj]
  {10.3847/1538-4357/aaadfb}, \href
  {https://ui.adsabs.harvard.edu/abs/2018ApJ...856...69D} {856, 69}

\bibitem[\protect\citeauthoryear{{Dekel} \& {Silk}}{{Dekel} \&
  {Silk}}{1986}]{Dekel1986}
{Dekel} A.,  {Silk} J.,  1986, \mn@doi [\apj] {10.1086/164050}, \href
  {https://ui.adsabs.harvard.edu/abs/1986ApJ...303...39D} {303, 39}

\bibitem[\protect\citeauthoryear{{Di Cintio}, {Mostoghiu}, {Knebe}  \&
  {Navarro}}{{Di Cintio} et~al.}{2021}]{DiCintio2021}
{Di Cintio} A.,  {Mostoghiu} R.,  {Knebe} A.,   {Navarro} J.~F.,  2021, \mn@doi
  [\mnras] {10.1093/mnras/stab1682}, \href
  {https://ui.adsabs.harvard.edu/abs/2021MNRAS.506..531D} {506, 531}

\bibitem[\protect\citeauthoryear{{Doumler}, {Hoffman}, {Courtois}  \&
  {Gottl{\"o}ber}}{{Doumler} et~al.}{2013}]{Doumler2013}
{Doumler} T.,  {Hoffman} Y.,  {Courtois} H.,   {Gottl{\"o}ber} S.,  2013,
  \mn@doi [\mnras] {10.1093/mnras/sts613}, \href
  {https://ui.adsabs.harvard.edu/abs/2013MNRAS.430..888D} {430, 888}

\bibitem[\protect\citeauthoryear{{Dutton} et~al.,}{{Dutton}
  et~al.}{2017}]{Dutton2017}
{Dutton} A.~A.,  et~al., 2017, \mn@doi [\mnras] {10.1093/mnras/stx458}, \href
  {https://ui.adsabs.harvard.edu/abs/2017MNRAS.467.4937D} {467, 4937}

\bibitem[\protect\citeauthoryear{{Einasto}, {Saar}, {Kaasik}  \&
  {Chernin}}{{Einasto} et~al.}{1974}]{Einasto1974}
{Einasto} J.,  {Saar} E.,  {Kaasik} A.,   {Chernin} A.~D.,  1974, \mn@doi
  [\nat] {10.1038/252111a0}, \href
  {https://ui.adsabs.harvard.edu/abs/1974Natur.252..111E} {252, 111}

\bibitem[\protect\citeauthoryear{{Escala} et~al.,}{{Escala}
  et~al.}{2018}]{Escala2018}
{Escala} I.,  et~al., 2018, \mn@doi [\mnras] {10.1093/mnras/stx2858}, \href
  {https://ui.adsabs.harvard.edu/abs/2018MNRAS.474.2194E} {474, 2194}

\bibitem[\protect\citeauthoryear{{Fillingham}, {Cooper}, {Pace},
  {Boylan-Kolchin}, {Bullock}, {Garrison-Kimmel}  \& {Wheeler}}{{Fillingham}
  et~al.}{2016}]{Fillingham2016}
{Fillingham} S.~P.,  {Cooper} M.~C.,  {Pace} A.~B.,  {Boylan-Kolchin} M.,
  {Bullock} J.~S.,  {Garrison-Kimmel} S.,   {Wheeler} C.,  2016, \mn@doi
  [\mnras] {10.1093/mnras/stw2131}, \href
  {https://ui.adsabs.harvard.edu/abs/2016MNRAS.463.1916F} {463, 1916}

\bibitem[\protect\citeauthoryear{{Fossati} et~al.,}{{Fossati}
  et~al.}{2015}]{Fossati2015}
{Fossati} M.,  et~al., 2015, \mn@doi [\mnras] {10.1093/mnras/stu2255}, \href
  {https://ui.adsabs.harvard.edu/abs/2015MNRAS.446.2582F} {446, 2582}

\bibitem[\protect\citeauthoryear{{Gallart} et~al.,}{{Gallart}
  et~al.}{2015}]{Gallart2015}
{Gallart} C.,  et~al., 2015, \mn@doi [\apjl] {10.1088/2041-8205/811/2/L18},
  \href {https://ui.adsabs.harvard.edu/abs/2015ApJ...811L..18G} {811, L18}

\bibitem[\protect\citeauthoryear{{Gallazzi}, {Charlot}, {Brinchmann}, {White}
  \& {Tremonti}}{{Gallazzi} et~al.}{2005}]{Gallazzi2005}
{Gallazzi} A.,  {Charlot} S.,  {Brinchmann} J.,  {White} S. D.~M.,   {Tremonti}
  C.~A.,  2005, \mn@doi [\mnras] {10.1111/j.1365-2966.2005.09321.x}, \href
  {https://ui.adsabs.harvard.edu/abs/2005MNRAS.362...41G} {362, 41}

\bibitem[\protect\citeauthoryear{{Garrison-Kimmel} et~al.,}{{Garrison-Kimmel}
  et~al.}{2019}]{Garrison2019}
{Garrison-Kimmel} S.,  et~al., 2019, \mn@doi [\mnras] {10.1093/mnras/stz2507},
  \href {https://ui.adsabs.harvard.edu/abs/2019MNRAS.489.4574G} {489, 4574}

\bibitem[\protect\citeauthoryear{{Genina}, {Frenk}, {Ben{\'\i}tez-Llambay},
  {Cole}, {Navarro}, {Oman}  \& {Fattahi}}{{Genina} et~al.}{2019}]{Genina2019}
{Genina} A.,  {Frenk} C.~S.,  {Ben{\'\i}tez-Llambay} A.~r.,  {Cole} S.,
  {Navarro} J.~F.,  {Oman} K.~A.,   {Fattahi} A.,  2019, \mn@doi [\mnras]
  {10.1093/mnras/stz1852}, \href
  {https://ui.adsabs.harvard.edu/abs/2019MNRAS.488.2312G} {488, 2312}

\bibitem[\protect\citeauthoryear{{Gottloeber}, {Hoffman}  \&
  {Yepes}}{{Gottloeber} et~al.}{2010}]{CLUES2010}
{Gottloeber} S.,  {Hoffman} Y.,   {Yepes} G.,  2010, arXiv e-prints, \href
  {https://ui.adsabs.harvard.edu/abs/2010arXiv1005.2687G} {p. arXiv:1005.2687}

\bibitem[\protect\citeauthoryear{{Grcevich} \& {Putman}}{{Grcevich} \&
  {Putman}}{2009}]{Grcevich2009}
{Grcevich} J.,  {Putman} M.~E.,  2009, \mn@doi [\apj]
  {10.1088/0004-637X/696/1/385}, \href
  {https://ui.adsabs.harvard.edu/abs/2009ApJ...696..385G} {696, 385}

\bibitem[\protect\citeauthoryear{{Gunn} \& {Gott}}{{Gunn} \&
  {Gott}}{1972}]{Gunn1972}
{Gunn} J.~E.,  {Gott} J.~Richard I.,  1972, \mn@doi [\apj] {10.1086/151605},
  \href {https://ui.adsabs.harvard.edu/abs/1972ApJ...176....1G} {176, 1}

\bibitem[\protect\citeauthoryear{Harris et~al.,}{Harris et~al.}{2020}]{numpy}
Harris C.~R.,  et~al., 2020, \mn@doi [Nature] {10.1038/s41586-020-2649-2}, 585,
  357

\bibitem[\protect\citeauthoryear{{Haynes} et~al.,}{{Haynes}
  et~al.}{2011}]{alfalfa}
{Haynes} M.~P.,  et~al., 2011, \mn@doi [\aj] {10.1088/0004-6256/142/5/170},
  \href {https://ui.adsabs.harvard.edu/abs/2011AJ....142..170H} {142, 170}

\bibitem[\protect\citeauthoryear{{Hester}}{{Hester}}{2006}]{Hester2006}
{Hester} J.~A.,  2006, \mn@doi [\apj] {10.1086/505614}, \href
  {https://ui.adsabs.harvard.edu/abs/2006ApJ...647..910H} {647, 910}

\bibitem[\protect\citeauthoryear{{Hoffman}}{{Hoffman}}{2009}]{Hoffman2009}
{Hoffman} Y.,  2009, {Gaussian Fields and Constrained Simulations of the
  Large-Scale Structure}.
pp 565--583, \mn@doi{10.1007/978-3-540-44767-2\_17}

\bibitem[\protect\citeauthoryear{{Hoffman} \& {Ribak}}{{Hoffman} \&
  {Ribak}}{1991}]{Hoffman1991}
{Hoffman} Y.,  {Ribak} E.,  1991, \mn@doi [\apjl] {10.1086/186160}, \href
  {https://ui.adsabs.harvard.edu/abs/1991ApJ...380L...5H} {380, L5}

\bibitem[\protect\citeauthoryear{{Hopkins} et~al.,}{{Hopkins}
  et~al.}{2018}]{Hopkins2018}
{Hopkins} P.~F.,  et~al., 2018, \mn@doi [\mnras] {10.1093/mnras/sty1690}, \href
  {https://ui.adsabs.harvard.edu/abs/2018MNRAS.480..800H} {480, 800}

\bibitem[\protect\citeauthoryear{Hunter}{Hunter}{2007}]{matplotlib}
Hunter J.~D.,  2007, \mn@doi [Computing in Science Engineering]
  {10.1109/MCSE.2007.55}, 9, 90

\bibitem[\protect\citeauthoryear{{Kauffmann} et~al.,}{{Kauffmann}
  et~al.}{2003}]{Kauffmann2003}
{Kauffmann} G.,  et~al., 2003, \mn@doi [\mnras]
  {10.1046/j.1365-8711.2003.06292.x}, \href
  {https://ui.adsabs.harvard.edu/abs/2003MNRAS.341...54K} {341, 54}

\bibitem[\protect\citeauthoryear{{Kawata} \& {Mulchaey}}{{Kawata} \&
  {Mulchaey}}{2008}]{Kawata2008}
{Kawata} D.,  {Mulchaey} J.~S.,  2008, \mn@doi [\apjl] {10.1086/526544}, \href
  {https://ui.adsabs.harvard.edu/abs/2008ApJ...672L.103K} {672, L103}

\bibitem[\protect\citeauthoryear{{Kawata}, {Gibson}, {Barnes}, {Grand}  \&
  {Rahimi}}{{Kawata} et~al.}{2014}]{Kawata2014}
{Kawata} D.,  {Gibson} B.~K.,  {Barnes} D.~J.,  {Grand} R. J.~J.,   {Rahimi}
  A.,  2014, \mn@doi [\mnras] {10.1093/mnras/stt2267}, \href
  {https://ui.adsabs.harvard.edu/abs/2014MNRAS.438.1208K} {438, 1208}

\bibitem[\protect\citeauthoryear{{Kennicutt}}{{Kennicutt}}{1998}]{Kennicutt1998}
{Kennicutt} Robert~C. J.,  1998, \mn@doi [\apj] {10.1086/305588}, \href
  {https://ui.adsabs.harvard.edu/abs/1998ApJ...498..541K} {498, 541}

\bibitem[\protect\citeauthoryear{{Kirby}, {Cohen}, {Guhathakurta}, {Cheng},
  {Bullock}  \& {Gallazzi}}{{Kirby} et~al.}{2013}]{Kirby2013}
{Kirby} E.~N.,  {Cohen} J.~G.,  {Guhathakurta} P.,  {Cheng} L.,  {Bullock}
  J.~S.,   {Gallazzi} A.,  2013, \mn@doi [\apj] {10.1088/0004-637X/779/2/102},
  \href {https://ui.adsabs.harvard.edu/abs/2013ApJ...779..102K} {779, 102}

\bibitem[\protect\citeauthoryear{{Kormendy} \& {Sanders}}{{Kormendy} \&
  {Sanders}}{1992}]{Kormendy1992}
{Kormendy} J.,  {Sanders} D.~B.,  1992, \mn@doi [\apjl] {10.1086/186370}, \href
  {https://ui.adsabs.harvard.edu/abs/1992ApJ...390L..53K} {390, L53}

\bibitem[\protect\citeauthoryear{{Kravtsov}, {Vikhlinin}  \&
  {Meshcheryakov}}{{Kravtsov} et~al.}{2018}]{Kravtsov2018}
{Kravtsov} A.~V.,  {Vikhlinin} A.~A.,   {Meshcheryakov} A.~V.,  2018, \mn@doi
  [Astronomy Letters] {10.1134/S1063773717120015}, \href
  {https://ui.adsabs.harvard.edu/abs/2018AstL...44....8K} {44, 8}

\bibitem[\protect\citeauthoryear{{Leroy}, {Walter}, {Brinks}, {Bigiel}, {de
  Blok}, {Madore}  \& {Thornley}}{{Leroy} et~al.}{2008}]{Leroy2008}
{Leroy} A.~K.,  {Walter} F.,  {Brinks} E.,  {Bigiel} F.,  {de Blok} W.~J.~G.,
  {Madore} B.,   {Thornley} M.~D.,  2008, \mn@doi [\aj]
  {10.1088/0004-6256/136/6/2782}, \href
  {https://ui.adsabs.harvard.edu/abs/2008AJ....136.2782L} {136, 2782}

\bibitem[\protect\citeauthoryear{{Libeskind} et~al.,}{{Libeskind}
  et~al.}{2020}]{Libeskind2020}
{Libeskind} N.~I.,  et~al., 2020, \mn@doi [\mnras] {10.1093/mnras/staa2541},
  \href {https://ui.adsabs.harvard.edu/abs/2020MNRAS.498.2968L} {498, 2968}

\bibitem[\protect\citeauthoryear{{Macci{\`o}}, {Udrescu}, {Dutton}, {Obreja},
  {Wang}, {Stinson}  \& {Kang}}{{Macci{\`o}} et~al.}{2016}]{Maccio2016}
{Macci{\`o}} A.~V.,  {Udrescu} S.~M.,  {Dutton} A.~A.,  {Obreja} A.,  {Wang}
  L.,  {Stinson} G.~R.,   {Kang} X.,  2016, \mn@doi [\mnras]
  {10.1093/mnrasl/slw147}, \href
  {https://ui.adsabs.harvard.edu/abs/2016MNRAS.463L..69M} {463, L69}

\bibitem[\protect\citeauthoryear{{Macci{\`o}}, {Frings}, {Buck}, {Penzo},
  {Dutton}, {Blank}  \& {Obreja}}{{Macci{\`o}} et~al.}{2017}]{Maccio2017}
{Macci{\`o}} A.~V.,  {Frings} J.,  {Buck} T.,  {Penzo} C.,  {Dutton} A.~A.,
  {Blank} M.,   {Obreja} A.,  2017, \mn@doi [\mnras] {10.1093/mnras/stx2048},
  \href {https://ui.adsabs.harvard.edu/abs/2017MNRAS.472.2356M} {472, 2356}

\bibitem[\protect\citeauthoryear{{Macci{\`o}}, {Frings}, {Buck}, {Dutton},
  {Blank}, {Obreja}  \& {Dixon}}{{Macci{\`o}} et~al.}{2019}]{Maccio2019}
{Macci{\`o}} A.~V.,  {Frings} J.,  {Buck} T.,  {Dutton} A.~A.,  {Blank} M.,
  {Obreja} A.,   {Dixon} K.~L.,  2019, \mn@doi [\mnras] {10.1093/mnras/stz327},
  \href {https://ui.adsabs.harvard.edu/abs/2019MNRAS.484.5400M} {484, 5400}

\bibitem[\protect\citeauthoryear{{Mateo}}{{Mateo}}{1998}]{Mateo1998}
{Mateo} M.~L.,  1998, \mn@doi [\araa] {10.1146/annurev.astro.36.1.435}, \href
  {https://ui.adsabs.harvard.edu/abs/1998ARA&A..36..435M} {36, 435}

\bibitem[\protect\citeauthoryear{{McConnachie}}{{McConnachie}}{2012}]{Mcconnachie2012}
{McConnachie} A.~W.,  2012, \mn@doi [\aj] {10.1088/0004-6256/144/1/4}, \href
  {https://ui.adsabs.harvard.edu/abs/2012AJ....144....4M} {144, 4}

\bibitem[\protect\citeauthoryear{{McGaugh}}{{McGaugh}}{2005}]{McGaugh2005}
{McGaugh} S.~S.,  2005, \mn@doi [\apj] {10.1086/432968}, \href
  {https://ui.adsabs.harvard.edu/abs/2005ApJ...632..859M} {632, 859}

\bibitem[\protect\citeauthoryear{{McGaugh}}{{McGaugh}}{2012}]{McGaugh2012}
{McGaugh} S.~S.,  2012, \mn@doi [\aj] {10.1088/0004-6256/143/2/40}, \href
  {https://ui.adsabs.harvard.edu/abs/2012AJ....143...40M} {143, 40}

\bibitem[\protect\citeauthoryear{{Moster}, {Macci\`o}  \&
  {Somerville}}{{Moster} et~al.}{2014}]{Moster2014}
{Moster} B.~P.,  {Macci\`o} A.~V.,   {Somerville} R.~S.,  2014, \mn@doi
  [\mnras] {10.1093/mnras/stt1702}, \href
  {https://ui.adsabs.harvard.edu/abs/2014MNRAS.437.1027M} {437, 1027}

\bibitem[\protect\citeauthoryear{{Obreja}, {Stinson}, {Dutton}, {Macci{\`o}},
  {Wang}  \& {Kang}}{{Obreja} et~al.}{2016}]{Obreja2016}
{Obreja} A.,  {Stinson} G.~S.,  {Dutton} A.~A.,  {Macci{\`o}} A.~V.,  {Wang}
  L.,   {Kang} X.,  2016, \mn@doi [\mnras] {10.1093/mnras/stw690}, \href
  {https://ui.adsabs.harvard.edu/abs/2016MNRAS.459..467O} {459, 467}

\bibitem[\protect\citeauthoryear{{Obreja} et~al.,}{{Obreja}
  et~al.}{2019}]{Obreja2019}
{Obreja} A.,  et~al., 2019, \mn@doi [\mnras] {10.1093/mnras/stz1563}, \href
  {https://ui.adsabs.harvard.edu/abs/2019MNRAS.487.4424O} {487, 4424}

\bibitem[\protect\citeauthoryear{{Oh} et~al.,}{{Oh} et~al.}{2015}]{Oh2015}
{Oh} S.-H.,  et~al., 2015, \mn@doi [\aj] {10.1088/0004-6256/149/6/180}, \href
  {https://ui.adsabs.harvard.edu/abs/2015AJ....149..180O} {149, 180}

\bibitem[\protect\citeauthoryear{{Okamoto}, {Arimoto}, {Yamada}  \&
  {Onodera}}{{Okamoto} et~al.}{2012}]{Okamoto2012}
{Okamoto} S.,  {Arimoto} N.,  {Yamada} Y.,   {Onodera} M.,  2012, \mn@doi
  [\apj] {10.1088/0004-637X/744/2/96}, \href
  {https://ui.adsabs.harvard.edu/abs/2012ApJ...744...96O} {744, 96}

\bibitem[\protect\citeauthoryear{{Peeples}, {Werk}, {Tumlinson}, {Oppenheimer},
  {Prochaska}, {Katz}  \& {Weinberg}}{{Peeples} et~al.}{2014}]{Peeples2014}
{Peeples} M.~S.,  {Werk} J.~K.,  {Tumlinson} J.,  {Oppenheimer} B.~D.,
  {Prochaska} J.~X.,  {Katz} N.,   {Weinberg} D.~H.,  2014, \mn@doi [\apj]
  {10.1088/0004-637X/786/1/54}, \href
  {https://ui.adsabs.harvard.edu/abs/2014ApJ...786...54P} {786, 54}

\bibitem[\protect\citeauthoryear{{Pilkington} et~al.,}{{Pilkington}
  et~al.}{2012}]{Pilkington2012}
{Pilkington} K.,  et~al., 2012, \mn@doi [\mnras]
  {10.1111/j.1365-2966.2012.21353.x}, \href
  {https://ui.adsabs.harvard.edu/abs/2012MNRAS.425..969P} {425, 969}

\bibitem[\protect\citeauthoryear{{Planck Collaboration} et~al.,}{{Planck
  Collaboration} et~al.}{2014}]{planck14}
{Planck Collaboration} et~al., 2014, \mn@doi [\aap]
  {10.1051/0004-6361/201321591}, \href
  {http://adsabs.harvard.edu/abs/2014A%26A...571A..16P} {571, A16}

\bibitem[\protect\citeauthoryear{{Pontzen}, {Ro{\v{s}}kar}, {Stinson}  \&
  {Woods}}{{Pontzen} et~al.}{2013}]{Pontzen2013}
{Pontzen} A.,  {Ro{\v{s}}kar} R.,  {Stinson} G.,   {Woods} R.,  2013, {pynbody:
  N-Body/SPH analysis for python} (\mn@eprint {ascl} {1305.002})

\bibitem[\protect\citeauthoryear{{Putman}, {Zheng}, {Price-Whelan}, {Grcevich},
  {Johnson}, {Tollerud}  \& {Peek}}{{Putman} et~al.}{2021}]{Putman2021}
{Putman} M.~E.,  {Zheng} Y.,  {Price-Whelan} A.~M.,  {Grcevich} J.,  {Johnson}
  A.~C.,  {Tollerud} E.,   {Peek} J. E.~G.,  2021, \mn@doi [\apj]
  {10.3847/1538-4357/abe391}, \href
  {https://ui.adsabs.harvard.edu/abs/2021ApJ...913...53P} {913, 53}

\bibitem[\protect\citeauthoryear{{Ritchie} \& {Thomas}}{{Ritchie} \&
  {Thomas}}{2001}]{Ritchie2001}
{Ritchie} B.~W.,  {Thomas} P.~A.,  2001, \mn@doi [\mnras]
  {10.1046/j.1365-8711.2001.04268.x}, \href
  {https://ui.adsabs.harvard.edu/abs/2001MNRAS.323..743R} {323, 743}

\bibitem[\protect\citeauthoryear{{Saintonge} et~al.,}{{Saintonge}
  et~al.}{2011}]{Saintonge2011}
{Saintonge} A.,  et~al., 2011, \mn@doi [\mnras]
  {10.1111/j.1365-2966.2011.18677.x}, \href
  {https://ui.adsabs.harvard.edu/abs/2011MNRAS.415...32S} {415, 32}

\bibitem[\protect\citeauthoryear{{Sand}, {Seth}, {Olszewski}, {Willman},
  {Zaritsky}  \& {Kallivayalil}}{{Sand} et~al.}{2010}]{Sand2010}
{Sand} D.~J.,  {Seth} A.,  {Olszewski} E.~W.,  {Willman} B.,  {Zaritsky} D.,
  {Kallivayalil} N.,  2010, \mn@doi [\apj] {10.1088/0004-637X/718/1/530}, \href
  {https://ui.adsabs.harvard.edu/abs/2010ApJ...718..530S} {718, 530}

\bibitem[\protect\citeauthoryear{{Sawala}, {Scannapieco}  \& {White}}{{Sawala}
  et~al.}{2012}]{Sawala2012}
{Sawala} T.,  {Scannapieco} C.,   {White} S.,  2012, \mn@doi [\mnras]
  {10.1111/j.1365-2966.2011.20181.x}, \href
  {https://ui.adsabs.harvard.edu/abs/2012MNRAS.420.1714S} {420, 1714}

\bibitem[\protect\citeauthoryear{{Shen}, {Wadsley}  \& {Stinson}}{{Shen}
  et~al.}{2010}]{Shen2010}
{Shen} S.,  {Wadsley} J.,   {Stinson} G.,  2010, \mn@doi [\mnras]
  {10.1111/j.1365-2966.2010.17047.x}, \href
  {https://ui.adsabs.harvard.edu/abs/2010MNRAS.407.1581S} {407, 1581}

\bibitem[\protect\citeauthoryear{{Simon} et~al.,}{{Simon}
  et~al.}{2021}]{Simon2021}
{Simon} J.~D.,  et~al., 2021, \mn@doi [\apj] {10.3847/1538-4357/abd31b}, \href
  {https://ui.adsabs.harvard.edu/abs/2021ApJ...908...18S} {908, 18}

\bibitem[\protect\citeauthoryear{{Sofue}}{{Sofue}}{2015}]{Sofue2015}
{Sofue} Y.,  2015, \mn@doi [\pasj] {10.1093/pasj/psv042}, \href
  {https://ui.adsabs.harvard.edu/abs/2015PASJ...67...75S} {67, 75}

\bibitem[\protect\citeauthoryear{{Sorce}}{{Sorce}}{2015}]{Sorce2015}
{Sorce} J.~G.,  2015, \mn@doi [\mnras] {10.1093/mnras/stv760}, \href
  {https://ui.adsabs.harvard.edu/abs/2015MNRAS.450.2644S} {450, 2644}

\bibitem[\protect\citeauthoryear{{Sorce}}{{Sorce}}{2018}]{Sorce2018}
{Sorce} J.~G.,  2018, \mn@doi [\mnras] {10.1093/mnras/sty1631}, \href
  {https://ui.adsabs.harvard.edu/abs/2018MNRAS.478.5199S} {478, 5199}

\bibitem[\protect\citeauthoryear{{Sorce} \& {Guo}}{{Sorce} \&
  {Guo}}{2016}]{Sorce2016}
{Sorce} J.~G.,  {Guo} Q.,  2016, \mn@doi [\mnras] {10.1093/mnras/stw341}, \href
  {https://ui.adsabs.harvard.edu/abs/2016MNRAS.458.2667S} {458, 2667}

\bibitem[\protect\citeauthoryear{{Sorce}, {Courtois}, {Gottl{\"o}ber},
  {Hoffman}  \& {Tully}}{{Sorce} et~al.}{2014}]{Sorce2014}
{Sorce} J.~G.,  {Courtois} H.~M.,  {Gottl{\"o}ber} S.,  {Hoffman} Y.,   {Tully}
  R.~B.,  2014, \mn@doi [\mnras] {10.1093/mnras/stt2153}, \href
  {https://ui.adsabs.harvard.edu/abs/2014MNRAS.437.3586S} {437, 3586}

\bibitem[\protect\citeauthoryear{{Sorce} et~al.,}{{Sorce}
  et~al.}{2016}]{Sorce2016b}
{Sorce} J.~G.,  et~al., 2016, \mn@doi [\mnras] {10.1093/mnras/stv2407}, \href
  {https://ui.adsabs.harvard.edu/abs/2016MNRAS.455.2078S} {455, 2078}

\bibitem[\protect\citeauthoryear{{Spekkens}, {Urbancic}, {Mason}, {Willman}  \&
  {Aguirre}}{{Spekkens} et~al.}{2014}]{Spekkens2014}
{Spekkens} K.,  {Urbancic} N.,  {Mason} B.~S.,  {Willman} B.,   {Aguirre}
  J.~E.,  2014, \mn@doi [\apjl] {10.1088/2041-8205/795/1/L5}, \href
  {https://ui.adsabs.harvard.edu/abs/2014ApJ...795L...5S} {795, L5}

\bibitem[\protect\citeauthoryear{{Stinson}, {Seth}, {Katz}, {Wadsley},
  {Governato}  \& {Quinn}}{{Stinson} et~al.}{2006}]{Stinson2006}
{Stinson} G.,  {Seth} A.,  {Katz} N.,  {Wadsley} J.,  {Governato} F.,   {Quinn}
  T.,  2006, \mn@doi [\mnras] {10.1111/j.1365-2966.2006.11097.x}, \href
  {https://ui.adsabs.harvard.edu/abs/2006MNRAS.373.1074S} {373, 1074}

\bibitem[\protect\citeauthoryear{{Stinson}, {Brook}, {Macci{\`o}}, {Wadsley},
  {Quinn}  \& {Couchman}}{{Stinson} et~al.}{2013}]{Stinson2013}
{Stinson} G.~S.,  {Brook} C.,  {Macci{\`o}} A.~V.,  {Wadsley} J.,  {Quinn}
  T.~R.,   {Couchman} H.~M.~P.,  2013, \mn@doi [\mnras] {10.1093/mnras/sts028},
  \href {https://ui.adsabs.harvard.edu/abs/2013MNRAS.428..129S} {428, 129}

\bibitem[\protect\citeauthoryear{{Su}, {Hopkins}, {Hayward},
  {Faucher-Gigu{\`e}re}, {Kere{\v{s}}}, {Ma}  \& {Robles}}{{Su}
  et~al.}{2017}]{Su2017}
{Su} K.-Y.,  {Hopkins} P.~F.,  {Hayward} C.~C.,  {Faucher-Gigu{\`e}re} C.-A.,
  {Kere{\v{s}}} D.,  {Ma} X.,   {Robles} V.~H.,  2017, \mn@doi [\mnras]
  {10.1093/mnras/stx1463}, \href
  {https://ui.adsabs.harvard.edu/abs/2017MNRAS.471..144S} {471, 144}

\bibitem[\protect\citeauthoryear{{Thomas}, {Maraston}, {Bender}  \& {Mendes de
  Oliveira}}{{Thomas} et~al.}{2005}]{Thomas2005}
{Thomas} D.,  {Maraston} C.,  {Bender} R.,   {Mendes de Oliveira} C.,  2005,
  \mn@doi [\apj] {10.1086/426932}, \href
  {https://ui.adsabs.harvard.edu/abs/2005ApJ...621..673T} {621, 673}

\bibitem[\protect\citeauthoryear{{Tully} et~al.,}{{Tully}
  et~al.}{2013}]{Tully2013}
{Tully} R.~B.,  et~al., 2013, \mn@doi [\aj] {10.1088/0004-6256/146/4/86}, \href
  {https://ui.adsabs.harvard.edu/abs/2013AJ....146...86T} {146, 86}

\bibitem[\protect\citeauthoryear{{Virtanen} et~al.,}{{Virtanen}
  et~al.}{2020}]{scipy}
{Virtanen} P.,  et~al., 2020, \mn@doi [Nature Methods]
  {10.1038/s41592-019-0686-2}, \href
  {https://ui.adsabs.harvard.edu/abs/2020NatMe..17..261V} {17, 261}

\bibitem[\protect\citeauthoryear{{Wadsley}, {Veeravalli}  \&
  {Couchman}}{{Wadsley} et~al.}{2008}]{Wadsley2008}
{Wadsley} J.~W.,  {Veeravalli} G.,   {Couchman} H.~M.~P.,  2008, \mn@doi
  [\mnras] {10.1111/j.1365-2966.2008.13260.x}, \href
  {https://ui.adsabs.harvard.edu/abs/2008MNRAS.387..427W} {387, 427}

\bibitem[\protect\citeauthoryear{{Wadsley}, {Keller}  \& {Quinn}}{{Wadsley}
  et~al.}{2017}]{Wadsley2017}
{Wadsley} J.~W.,  {Keller} B.~W.,   {Quinn} T.~R.,  2017, \mn@doi [\mnras]
  {10.1093/mnras/stx1643}, \href
  {https://ui.adsabs.harvard.edu/abs/2017MNRAS.471.2357W} {471, 2357}

\bibitem[\protect\citeauthoryear{{Wang}, {Dutton}, {Stinson}, {Macci\`o},
  {Penzo}, {Kang}, {Keller}  \& {Wadsley}}{{Wang} et~al.}{2015}]{nihao_main}
{Wang} L.,  {Dutton} A.~A.,  {Stinson} G.~S.,  {Macci\`o} A.~V.,  {Penzo} C.,
  {Kang} X.,  {Keller} B.~W.,   {Wadsley} J.,  2015, \mn@doi [\mnras]
  {10.1093/mnras/stv1937}, \href
  {https://ui.adsabs.harvard.edu/abs/2015MNRAS.454...83W} {454, 83}

\bibitem[\protect\citeauthoryear{{Weisz}, {Dolphin}, {Skillman}, {Holtzman},
  {Gilbert}, {Dalcanton}  \& {Williams}}{{Weisz} et~al.}{2014}]{Weisz2014}
{Weisz} D.~R.,  {Dolphin} A.~E.,  {Skillman} E.~D.,  {Holtzman} J.,  {Gilbert}
  K.~M.,  {Dalcanton} J.~J.,   {Williams} B.~F.,  2014, \mn@doi [\apj]
  {10.1088/0004-637X/789/2/148}, \href
  {https://ui.adsabs.harvard.edu/abs/2014ApJ...789..148W} {789, 148}

\bibitem[\protect\citeauthoryear{{White} \& {Frenk}}{{White} \&
  {Frenk}}{1991}]{White1991}
{White} S. D.~M.,  {Frenk} C.~S.,  1991, \mn@doi [\apj] {10.1086/170483}, \href
  {https://ui.adsabs.harvard.edu/abs/1991ApJ...379...52W} {379, 52}

\bibitem[\protect\citeauthoryear{{White} \& {Rees}}{{White} \&
  {Rees}}{1978}]{White1978}
{White} S.~D.~M.,  {Rees} M.~J.,  1978, \mn@doi [\mnras]
  {10.1093/mnras/183.3.341}, \href
  {https://ui.adsabs.harvard.edu/abs/1978MNRAS.183..341W} {183, 341}

\bibitem[\protect\citeauthoryear{{Williamson}, {Martel}  \&
  {Romeo}}{{Williamson} et~al.}{2016}]{Williamson2016}
{Williamson} D.,  {Martel} H.,   {Romeo} A.~B.,  2016, \mn@doi [\apj]
  {10.3847/0004-637X/831/1/1}, \href
  {https://ui.adsabs.harvard.edu/abs/2016ApJ...831....1W} {831, 1}

\bibitem[\protect\citeauthoryear{{Wilman}, {Zibetti}  \&
  {Budav{\'a}ri}}{{Wilman} et~al.}{2010}]{Wilman2010}
{Wilman} D.~J.,  {Zibetti} S.,   {Budav{\'a}ri} T.,  2010, \mn@doi [\mnras]
  {10.1111/j.1365-2966.2010.16845.x}, \href
  {https://ui.adsabs.harvard.edu/abs/2010MNRAS.406.1701W} {406, 1701}

\bibitem[\protect\citeauthoryear{{Yin}, {Hou}, {Prantzos}, {Boissier}, {Chang},
  {Shen}  \& {Zhang}}{{Yin} et~al.}{2009}]{Yin2009}
{Yin} J.,  {Hou} J.~L.,  {Prantzos} N.,  {Boissier} S.,  {Chang} R.~X.,  {Shen}
  S.~Y.,   {Zhang} B.,  2009, \mn@doi [\aap] {10.1051/0004-6361/200912316},
  \href {https://ui.adsabs.harvard.edu/abs/2009A&A...505..497Y} {505, 497}

\bibitem[\protect\citeauthoryear{{Zaroubi}, {Hoffman}, {Fisher}  \&
  {Lahav}}{{Zaroubi} et~al.}{1995}]{Zaroubi1995}
{Zaroubi} S.,  {Hoffman} Y.,  {Fisher} K.~B.,   {Lahav} O.,  1995, \mn@doi
  [\apj] {10.1086/176070}, \href
  {https://ui.adsabs.harvard.edu/abs/1995ApJ...449..446Z} {449, 446}

\makeatother
\end{thebibliography}



\appendix

\section{NIHAO-LG(nmd)}\label{sec:nmd}

\begin{figure*}
    \centering
    \includegraphics[width=\linewidth]{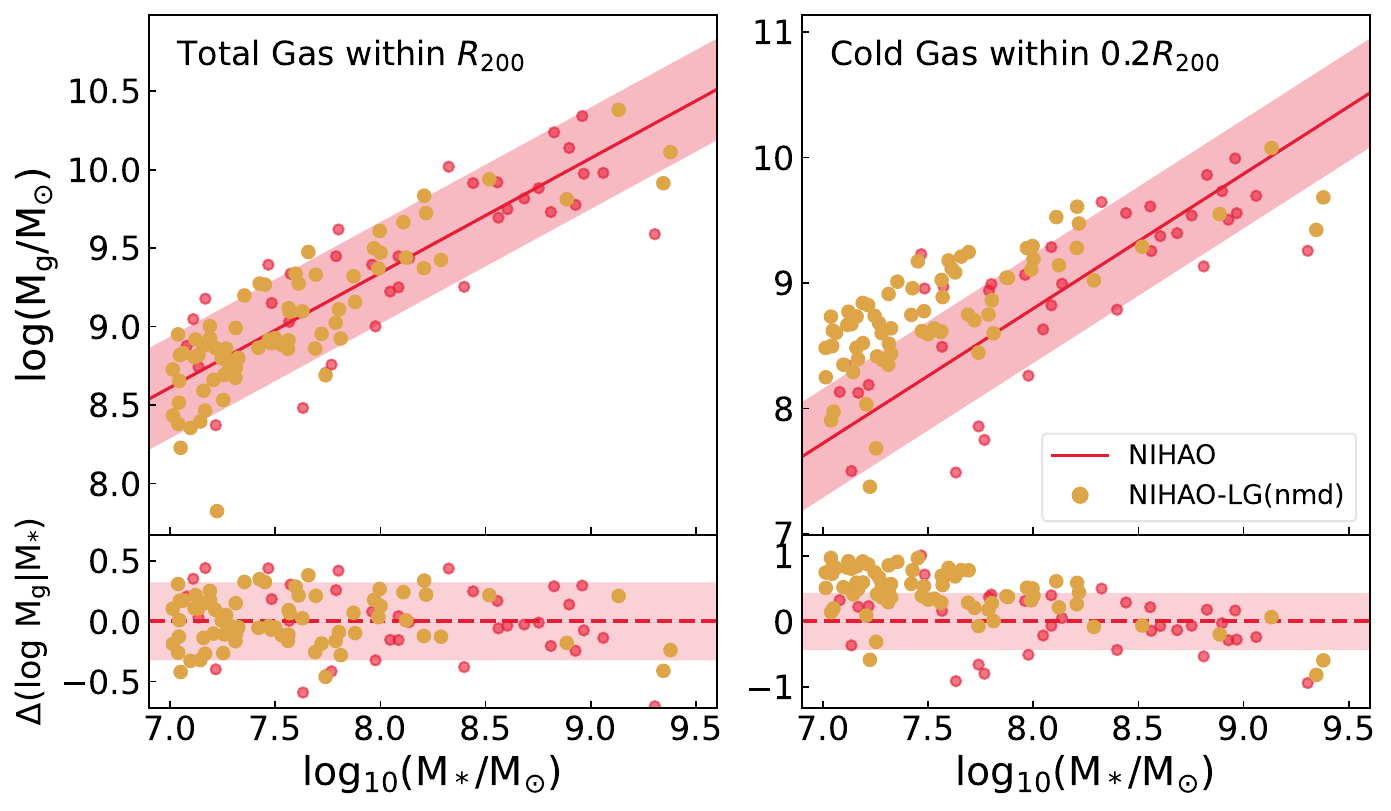}
    \caption{Total gas mass (left-hand panel) and cold gas mass ($T<20000\,{\rm K}$; right-hand panel) versus stellar mass at redshift \rz for NIHAO field (red circles) and NIHAO-LG(nmd) presented as gold circles dwarf galaxies. 
    The solid red line and shaded region represent a best fit of the NIHAO field dwarf galaxies and 1$\sigma$ scatter about that fit, respectively. 
    Total gas masses were measured within $\rm R_{200}$, while cold gas and stellar masses were measured within $\rm 0.2R_{200}$. 
    The residuals with respect to the NIHAO field dwarf best fit are shown in both bottom panels.}
    \label{fig:gasmass_nmd}
\end{figure*}

\begin{figure*}
    \centering
    \includegraphics[width=0.32\textwidth]{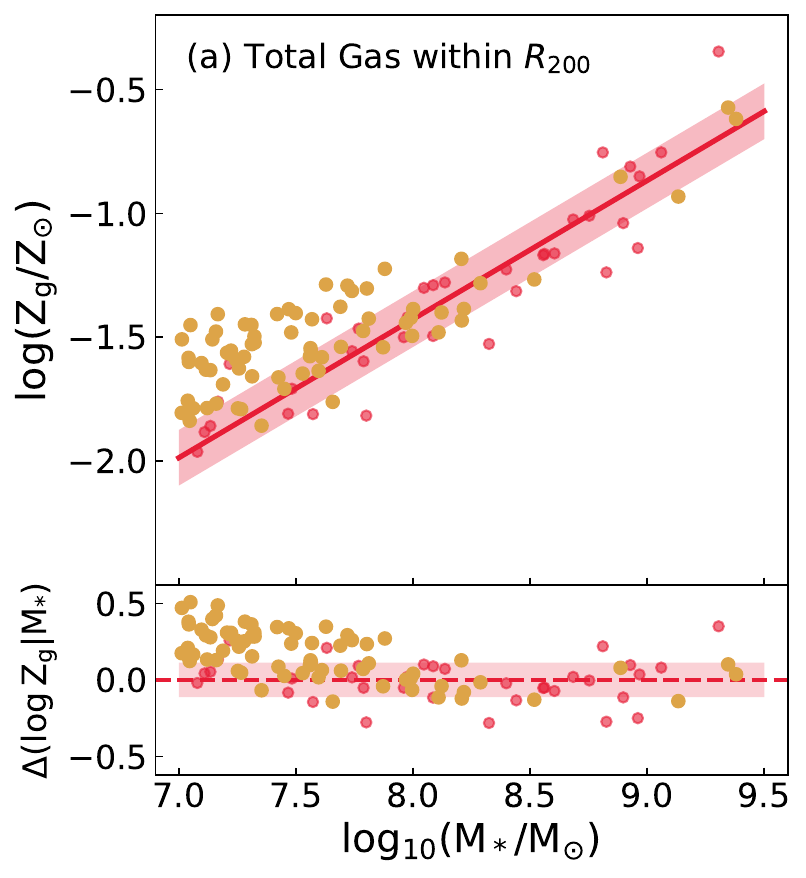}
    \includegraphics[width=0.32\textwidth]{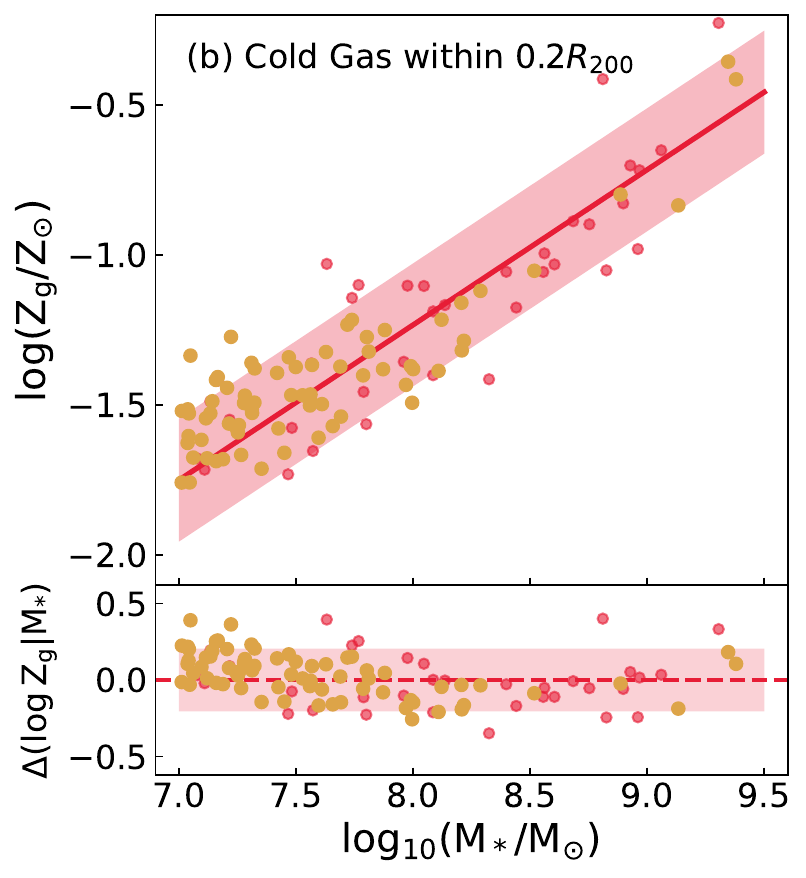}
    \includegraphics[width=0.32\textwidth]{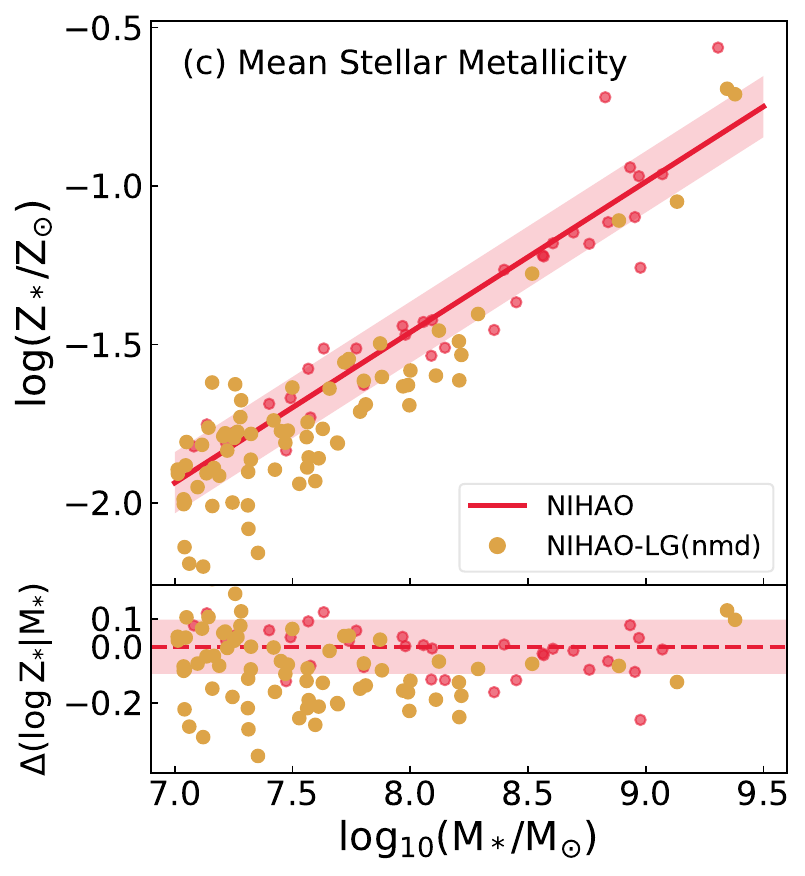}
    \caption{Dwarf galaxy comparison between NIHAO field and NIHAO-LG(nmd) for various properties related to metal content. 
    Panel a shows the total gas metallicity within R$_{\rm 200}$, panel b shows the cold gas metallicity within 0.2R$_{\rm 200}$, and panel c shows the mean stellar metallicity. 
    The red dots, lines, and shaded regions present the NIHAO field galaxies, linear fit, and scatter, respectively. 
    The gold points show dwarfs from the NIHAO-LG(nmd) simulation.}
    \label{fig:metal_nmd}
\end{figure*}

Figs. \ref{fig:gasmass_nmd} and \ref{fig:metal_nmd} reproduce the NIHAO field and NIHAO-LG(nmd) comparisons for various galaxy properties.
At redshift \rz, we present different measures of gas masses (total gas within R$_{\rm 200}$ and cold gas within 0.2R$_{\rm 200}$) as well as various measure of metal contents (average gas metallicity within R$_{\rm 200}$, average cold gas metallicity 0.2R$_{\rm 200}$, and mean stellar metallicity). 
Similar to NIHAO-LG, we find that the NIHAO-LG(nmd) dwarfs show excess cold gas within 0.2R$_{\rm 200}$ and total gas-phase metals relative to NIHAO field dwarfs.  
Independent of the constrained LG simulations used, our conclusions are unchanged.

\section{Iron Abundance in the LG} \label{sec:iron}

\begin{figure*}
    \centering
    \includegraphics[width=\linewidth]{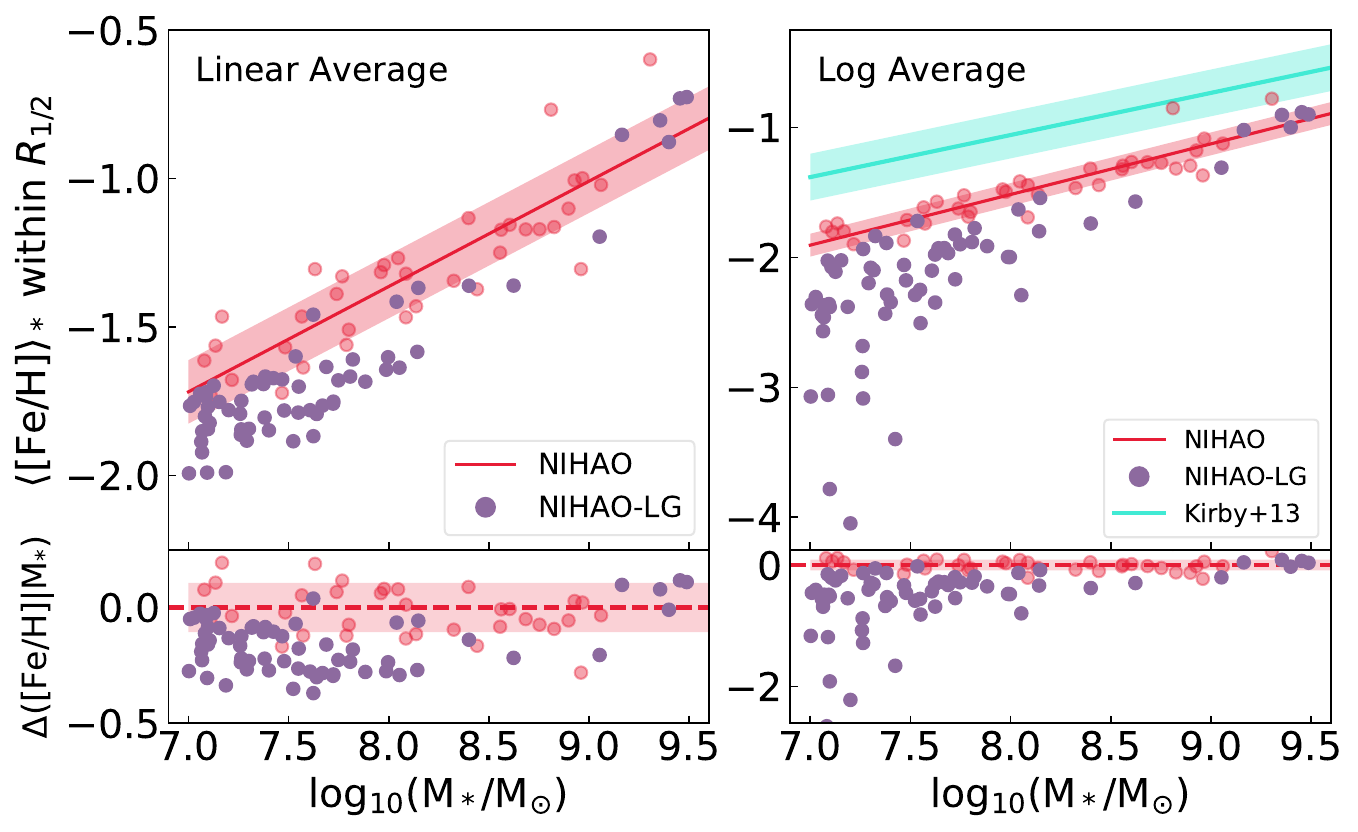}
    \caption{Iron abundance, $\langle [{\rm Fe/H}]\rangle_*$, versus stellar mass for NIHAO field and NIHAO-LG dwarfs.
    The color scheme is as in \Fig{gasmass}. 
    $\langle [{\rm Fe/H}]\rangle_*$ was averaged over all stellar particles within the half-light radius, $R_{\rm 1/2}$. 
    The left-hand panel shows mass-weighted averages carried out in linear space (see \Eq{linfeh} for definition) while the right-hand panel shows mass-weighted averages calculated in log space as presented in \Eq{logfeh}. 
    In cyan, the right-hand panel shows the observed relation for LG dwarf galaxies from \protect\cite{Kirby2013}.}
    \label{fig:feh}
\end{figure*}

Observers typically measure stellar chemical abundance of galaxies using bright metal absorption lines such as those of iron or magnesium. 
Here we present a comparison of the stellar iron abundances between NIHAO field and NIHAO-LG dwarfs and observations from \cite{Kirby2013}.
With most of the iron concentrated in the central parts of the galaxies, the average iron abundance weighted by stellar mass encompasses all star particles within the half-light radius, $R_{\rm 1/2}$.
An essential aspect of this chemical comparison is the averaging process for the star particles in the simulation. 
At first, the averaging is carried out with the metallicity in logarithmic space, also called the ``geometric mean'' (hereafter, log average; Kirby, private communication).
The average iron abundance then becomes,

\begin{equation}
    \langle{\rm [Fe/H]}\rangle_{\rm *,log} = \frac{\displaystyle\sum_i ({\rm [Fe/H]}_{*,i} m_{*,i})}{\displaystyle\sum_i(m_{*, i})},
    \label{eq:logfeh}
\end{equation}

\noindent where $\rm [Fe/H]$ is the logarithmic iron abundance and $m_{\rm *}$ is the stellar mass of the star particle $i$.
However, the averaging process in logarithmic units is inappropriate. 
A proper expression for the mean iron abundance is given by

\begin{equation}
    \langle{\rm [Fe/H]}\rangle_{\rm *,lin} = \log \left( \frac{\displaystyle\sum_i (10^{{\rm [Fe/H]}_{*,i}}m_{*,i})}{\displaystyle\sum_i(m_{*, i})} \right).
    \label{eq:linfeh} 
\end{equation}
In \Eq{linfeh}, the iron abundance per star particle, $i$, is calculated in linear space and the logarithm of the complete expression is taken once the average is computed.
Our comparison of the simulated dwarf systems uses linear averages presented in \Eq{linfeh}. 
For complementarity and comparison with observations, the stellar iron abundances averaged in logarithmic space are also reported.

The left-hand panel of \Fig{feh} presents the linearly averaged iron abundance versus stellar mass for NIHAO field and NIHAO-LG dwarf samples. 
Field dwarf galaxies from NIHAO follow $\langle{\rm [Fe/H]}\rangle\propto M_{\rm *}^{0.36\pm0.04}$ with a scatter of $0.09\pm0.02$ dex.
\nlg low-mass dwarfs are somewhat poorer in metals than the field.
A linear average for the stellar iron abundance is preferentially biased towards metal-rich stars.

\nlg shows metal-poor stellar populations relative to field systems.
Given the deep potential of the MW and M31, a significant fraction of the outflowing gas and metals may be retained by the massive halo through galactic fountain effect.
With most of the metal retained by the massive halos, the LG dwarf populations would show a metal poorer stellar population relative to expectations due to pre-enrichment.
Our result in the left-hand panel of \Fig{feh} is quantitatively similar to that of FIRE simulations \citep{Escala2018} or the high-resolution NIHAO simulations \citep{Buck2019}.

The right-hand panel of \Fig{feh} shows the logarithmic averaged stellar iron abundance for the NIHAO field and NIHAO-LG dwarfs and observed relation from \cite{Kirby2013}.
With a different averaging technique comes a revised relation;
the field galaxies relation, $\langle{\rm [Fe/H]}\rangle_*$--$M_{\rm *}$, is now slightly steeper and tighter with a slope of $0.39\pm0.04$ and scatter $0.09\pm0.02$.
The differences between linear and logarithmic averaging of the $\langle{\rm [Fe/H]}\rangle_*$--$M_{\rm *}$ relations are still well within the confidence intervals. 
The observed relation shows similar linear slope ($0.30\pm 0.02$) but with a larger scatter (0.17 dex) and lower zero-point. 
The simulated field galaxies from NIHAO are more iron-poor, as is expected for these low-mass galaxies which cannot retain metal-rich gas due to stellar feedback and winds.
The NIHAO-LG dwarfs show an even poorer metal content in stars, with their [Fe/H] distribution lower than observations by a factor 15. 
The simulations and observations differ only in their zero-point offset (the trends have the same slope) which is directly linked to the implementation of supernova feedback in the simulations \citep{Escala2018}.
The study of \cite{Buck2021} provides an improved chemical enrichment scheme for NIHAO simulations that should yield a better match with observations.
We shall return to such data-model comparison in a future study.

\bsp	
\label{lastpage}
\end{document}